# Technical Analysis and Discrete False Discovery Rate: Evidence from MSCI Indices

Georgios Sermpinis[*]*, Arman Hassanniakalager[†], Charalampos Stasinakis[‡], Ioannis Psaradellis [§]

**Version:** 10 June 2019

## Abstract

We investigate the performance of dynamic portfolios constructed using more than 21,000 technical trading rules on 12 categorical and country-specific markets over the 2004-2015 study period, on rolling forward structures of different lengths. We also introduce a discrete false discovery rate (DFRD[+/-]) method for controlling data snooping bias. Compared to the existing methods, DFRD[+/-] is adaptive and more powerful, and accommodates for discrete *p*-values. The profitability, persistence and robustness of the technical rules are examined. Technical analysis still has short-term value in advanced, emerging and frontier markets. Financial stress, the economic environment and market development seem to affect the performance of trading rules. A cross-validation exercise highlights the importance of frequent rebalancing and the variability of profitability in trading with technical analysis.

[*]*Corresponding author: Adam Smith Business School, University of Glasgow, Glasgow, G12 8QQ, UK
[†] School of Management, University of Bath, Bath, UK
[‡] Adam Smith Business School, University of Glasgow, Glasgow, UK
[§] School of Economics & Finance, University of St Andrews, St Andrews, UK

# 1. Introduction

Technical analysis constitutes the type of investment analysis which uses simple mathematical formulations or graphical representations of financial assets' time series to explore trading opportunities. In its algorithmic, and thus more quantitative form, it utilizes the analysis of the asset price's history, volume data and summary statistics, through mathematical tools, usually referred to as technical indicators and oscillators. Even though this form of analysis has been widely exploited by both investors and academics over the years, there is a long and ongoing discussion about whether it truly has predictive power and can generate significant profitability in equities markets. Previous literature is split into studies highlighting the genuine profitability of technical analysis (see among others, Brock *et al.,* 1992; Hsu *et al.,* 2010) and those arguing against it (see among others, Sullivan *et al.,* 1999; Bajgrowicz and Scaillet, 2012).

The success of technical trading rules is associated with asset pricing anomalies, such as momentum (Jegadeesh, 1993; Assness, 2013) and reversal (DeBond and Thaler, 1985; Jegadeesh, 1990). Momentum is a result of initial under-reaction or delayed over-reaction of investors to securities' price movements, while reversal is a result of initial over-reaction. Some recent studies have tried to use this information and employ technical trading rules for the creation of universal trend factors and the optimization of asset allocation in portfolio construction (see Zhu and Zhou, 2009; Han *et al.,* 2016). From the point of view of practitioners, the use of technical analysis is not that debatable, and many successful hedge funds and portfolio managers make substantial use of technical trading (see Fung and Hsieh, 2001; Lo and Hasanhodzic, 2009).

Nonetheless a comprehensive and up-to-date analysis of technical trading on equity markets is still being asked for by both academics and practitioners. This is because the majority of previous studies tend to be narrowly focused on specific aspects of technical analysis of equity indices. For example, looking at a single market index, a restricted number and classes of technical trading rules, no disaggregated analysis of the classes that are profitable, or a "sterilized" exercise of technical analysis (e.g. no transaction costs involved) is totally different from the way traders operate in practice. Other issues, such as the size of the in-sample (IS)/out-of-sample (OOS) or the level of financial stress and market development are also frequently ignored. In addition, no study has consistently measured the persistence of technical analysis, the role of frequency in portfolio rebalancing or whether the identified rules survive backtesting.

Testing a large set of technical trading rules on a given data set risks false rejections of the null of efficiency, as some rules in a large set will, by chance, prove to be profitable ex post. A researcher first needs to adopt a multiple hypothesis testing (MHT) framework in order to identify whether truly profitable rules exist. The framework should be powerful, adaptive and computationally efficient. Even if a group of technical rules demonstrates significant excess profitability in specific markets, there are still several questions which need to be answered. For example, how long does this profitability persist, are the markets during these periods under stress or turmoil, what are the optimal IS/OOS ratios for achieving the best performance or at what level of transaction costs is excess profitability neutralized? Another important issue which also needs to be investigated further is whether the level of financial stress and the level of market development lead to short-term anomalies in market efficiency, which allow for such profitability.

For the above reasons, we conduct one of the most extensive studies of technical trading in the equities markets ever carried out, as well as proposing a new method controlling for data snooping bias while adjusting for the potential issues found in previous techniques. We investigate the daily data of nine individual Morgan Stanley Capital International (MSCI) indices and three general MSCI indices replicating the performance of Advanced, Emerging and Frontier markets covering the period from 2004 to 2015. As for the technical trading rules considered, we employ the expanded universe of 21,000 rules of Hsu, Taylor and Wang (2016) incorporating several classes of technical trading indicators and oscillators. Our novel methodology accounting for data snooping is based on the *false discovery rate* (FDR) criterion and expands the $FDR^{+/-}$ approach of Barras *et al.* (2010) in numerous ways. Their $FDR^{+/-}$ detects a sufficiently large number of statistically significant positive rules, while allowing for a small number of false discoveries (Barras *et al.*, 2010; Brajgowicz and Scaillet, 2012). However, when usual resampling procedures (e.g. bootstrapping) are employed to compute each rule's corresponding *p*-values, large-scale homogeneous discrete *p*-values are realized, rather than the uniformly distributed continuous ones adopted in all previous studies (Storey, 2002; Storey *et al.,* 2004; Barras *et al.,* 2010; Brajgowicz and Scaillet, 2012). In addition, previous approaches can lead to unnecessary conservativeness and hence, poor estimations of the proportion of rules with no significant performance during the parameterization stage of $FDR^{+/-}$. In contrast, our discrete false discovery rate approach ($DFRD^{+/-}$) circumvents these issues by constructing a large-scale homogeneous discrete *p*-values framework, with its main parameters being dynamically estimated. Hence, we provide a fully adaptive, computationally and efficient approach controlling for data snooping bias, which can provide assistance to investors when it comes to OOS estimation, portfolio construction and improved decision-making in general. Moreover, our proposed method identifies performance persistence of successful portfolios of technical trading rules up to three months for some markets (i.e. emerging and frontier), which signifies that it is a potentially powerful tool for hedge fund and portfolio managers.

We employ our proposed method to perform several robustness checks and provide answers to the questions raised earlier related to break-even transaction costs and OOS estimation. Specifically, a break-even analysis of transaction costs of the outperforming trading rules is carried out, while we investigate the OOS performance of the rules in a rolling-forward structure, just as fund managers do in practice. Moreover, we analyse their performance persistence during the same periods as well as providing evidence of their performance during periods of turmoil, using stress indices provided by the Office of Financial Research (OFR). Finally, we introduce an innovative method for selecting the significant technical trading rules by cross-validating their performance between the full sample and IS/OOS.

Regarding our results, we find evidence of technical rules profitability after transaction costs in all markets studied, with trend-following families, such as moving averages and channel breakouts to dominate the contrarian ones. The profitability is stronger in emerging and frontier markets compared to the advanced ones. The persistence of this profitability varies over the years but is weak in general. Financial stress seems to have an effect on the predictability and excess profitability of technical analysis. More specifically, the profitability of technical analysis is 5-6 times higher for the US when financial stress levels are low, while the opposite trend is observed for other advanced markets and emerging ones.

The rest of this work is laid out as follows. Section 2 reports the examined data set and demonstrates the technical trading rules as well as the performance metrics employed in this study. Section 3 presents a discussion on data snooping bias testing as well as the relevant literature. The proposed methodology is thoroughly explained in Section 4. The empirical results for the IS performance are displayed in Section 5, while those for the OOS are presented in Section 6. Section 7 provides our concluding remarks.

## 2. Data set, Technical Trading Rules and Performance Metrics

This section presents the details of the setting at which the DFDR$^{+/-}$ is applied. In Section 2.1 the data set with the information regarding the examined markets is introduced. Section 2.2 describes the technical rules universe and Section 2.3 covers the performance measures used to compare the trading rules for different markets.

### 2.1 Data set

In this paper, we study nine Morgan Stanley Capital International (MSCI) indices that replicate the performance of markets in the United States, United Kingdom, Japan, Brazil, China, Russia, Estonia, Jordan and Morocco and the three general MSCI indices that replicate the World (Developed), Emerging and Frontier markets in total. The MSCI indices are market cap weighted indices that reflect the holding returns of US investors in different markets. They are denoted in US dollars and are important references for institutional investor

[1] ( see among others, Hsu, 2010; Bena *et al.,* 2017). They include large and mid-cap segments of the benchmark markets and thus mitigate liquidity and tradability issues. The sample period for all time series starts on 1 January 2004 and ends on 31 December 2016. The summary statistics of the log returns of the twelve series and of the risk-free rates series are presented in Table 1.

[Table 1 here]

All indices are leptokurtic while the risk-free rate series behaves very close to the normal curve. The UK, Brazil and Russia display quite high kurtosis. All time series except for the Frontier index and the risk-free rate exhibit negative skewness (with the UK having the least). The positive autocorrelation coefficient is seen for all times series except for Japan and the US, but the reported coefficient is not statistically significant for Japan.

### 2.2 Technical Rules

The purpose of technical trading strategies is to generate long (short) positions for the coming period based on historical quotes for open, high, low and closing prices along with other characteristics such as previous trends/momentums and directional movements. As already mentioned, their efficacy mainly lies with popular market price anomalies (i.e. momentum and reversals). In this study, 21,195 technical trading rules are utilized following the work of Hsu *et al.* (2016) for each of the twelve MSCI indices under study. This universe mainly includes trend-following rules such as Filter Rules (FRs), Moving Averages (MAs), Support-Resistance rules (SRs) and Channel Breakout rules (CBs), and a class of contrarian rules such as the Relative Strength Indicators (RSIs). All of these use lag lengths representing specific components (e.g. daily, weekly, monthly, annual) of price trends and reversals that make sense economically. These technical indicators are commonly practised by trading desks and hedge funds, while they are available on finance websites, and in numerous finance research papers and textbooks[2].

---

[1] 99% of the top global investment managers apply MSCI indices (see P&I AUM data and MSCI clients as of December 2017).

The scope of our empirical analysis in Sections 5 and 6 is to explore the value of technical analysis in trading through our $DFDR^{+/-}$ procedure. Thus, we explore a very large and diverse universe of technical rules, an approach common to all the related literature that explores thousands of technical rules (see among others, Brock *et al.,* 1992; Hsu *et al.,* 2010; Sullivan *et al.,* 1999; Bajgrowicz and Scaillet, 2012; Hsu *et al.,* 2016). Naturally, some technical rules specifications will have more economic plausibility than others. Thus, it makes sense to vary the size of the test statistic($\varphi_j$) based on how much economic sense each technical rule makes (Harvey, 2017). However, the application of this approach should lead to modifications in both the bootstrap and FDR control approach. To the best of our knowledge, no similar procedure has been presented in the literature, something that can certainly present an interesting research direction.

Filter Rules (FRs): *are simple momentum techniques that generate buy (sell) signals, when the market price rises (drops) by more than a given percentage from its previous low (high). Setting the threshold percentage allows investors not to be misled by small market price fluctuations.*

Relative Strength Indicators (RSIs): *are contrarian oscillators that measure the speed of change of rapid price movements towards the mean, through overbought/oversold levels. Mean reversion is measured as the ratio of higher closes to lower closes (i.e. stocks with more or stronger positive changes have a higher RSI than stocks with the same level of negative changes). The stock is considered overbought when the RSI is above 70 and oversold when it*

---

[2] Earlier studies on technical analysis profitability include those of Levy (1967, 1971), Wilder (1978), Allen and Taylor (1992) and Taylor and Allen (1992).

*falls below 30 as the level of upward movements relative to the downward ones is usually normalized between 0 and 100.*

Moving Averages (MAs): *Technical analysts explore simple MAs or combinations of them. For example, uptrends start to form when the daily price MA exceeds the MA at a specific percent. The general rule is that long (short) positions are kept as long as the price remains above (below) the MA benchmark. Apart from their simplest version, double and triple MA rules are also considered including fast-slow variations along with the rest of the parameterizations.*

Support-Resistance rules (SRs): *trading rules are based on the premise that the price should remain in a trading range capped by a resistance and floored by a support level. Breaching these levels suggests that a stock will keep moving towards that direction and initiate momentum. A resistance-and-support level is predefined as the highest and lowest closing price of the previous -d closing prices respectively where -d is chosen by the trader.*

Channel Breakout rules (CBs): *similar to time-varying SRs, are parallel trends lines that form a trading channel. A signal is initiated when the price breaks either the upper or the lower bound of the channel. CBs perform well in markets revealing strong trends.*

For the exact characteristics of our technical rules, we refer the reader to Appendix A in Hsu *et al.* (2016).

## 2.3 Excess Returns, Transaction Costs and Performance Metrics

In this section, we define the daily return for every index examined as well as the performance metrics employed after accounting for transaction costs. First, we calculate the daily gross return from buying and holding the index during the prediction period:

$$r_t = \ln\left(\frac{P_t}{P_{t-1}}\right)$$

where $P_t$ is the spot price on day $t$, and $P_{t-1}$ is the spot price on the previous day. Each calendar year in our database has on average 260 trading days.

Secondly, we need to consider the impact of transaction costs on the technical trading simulation. For that reason, we treat transaction costs "*endogenously*" to the trading process. For instance, we deduct one-way transaction costs every time a long or short position is closed according to the next period's index value prediction. We estimate the one-way transaction cost taken at time $t$ for trading rule $j$ as:

$$TC_{j,t} = I_{j,t} \times tc \times P_t$$

where, $I_{j,t}$ is the indicator set to 1 when a position is closed for the studied trading rule and the transaction cost $TC_{j,t}$ is deducted (0 otherwise) at time *t* and *tc* represents the level of transaction costs used in basis points (bps).

The transaction cost can negatively affect the performance of the portfolios (Cesari and Cremonini, 2003). Industry-based factsheets, traders, retail online brokers, hedge fund managers and the academic literature recommend a transaction cost of 25-50 bps for trading MSCI indices (see for example, Cesari and Cremonini, 2003; Investment Technology Group, 2013; Eurex, 2018). MSCI (2013) suggests transaction costs up to 50 bps for their indices. In this study, we consider a one-way proportional transaction cost of 25 bps for advanced markets (US, UK, Japan, and Developed) and 50 bps for the other markets. These costs correspond to fees, bid-ask spread and slippage. These costs are realistic for large institutional investors.

In terms of performance metrics, we provide the annualized *mean excess return* and *Sharpe ratio*. In this way, we consider an absolute measure based on each technical trading rule's returns, such as the mean excess return, and a relative performance measure reporting the ratio of the mean excess return to the total risk of the investment in terms of excess returns' standard deviation (the Sharpe ratio). Denoting the trading signal triggered from a trading rule $j$, $1 \leq j \leq l$ (where = 21,195) at the end of each prediction period $t - 1$ ($\tau \leq t \leq T$) as $s_{j,t-1}$, where $s_{j,t-1} = 1, 0,$ or $-1$ represents a long, neutral or short position taken at time *t*, the mean excess return criterion $\overline{f}_{j,t}$ for the trading rule *j* is given by:

$$\overline{f}_{j,t} = \frac{1}{N}\sum_{t=\tau}^{T}[s_{j,t-1}r_t - TC_{j,t} - \ln(1 + r_{f,t})], \qquad j = 1, \dots, l$$

where $N = T - \tau + 1$ is the number of days examined and $r_{f,t}$ is the risk-free rate at time *t*. The $\tau$ is the activation period, since some of the technical trading rules use lagged values of indices up to one year (260 days). For the risk-free rate, we use the effective federal funds rate reported by the Federal Reserve in the US. Since the quotes for the risk-free rate are reported on an annual basis, we transform the rates into the daily values by $r_{f,t} = (1 + S_t)^{\frac{1}{260}} - 1$, where $r_{f,t}$ is the estimated daily rate and $S_t$ is the quoted federal funds rate.

Then, the Sharpe ratio metric expression $SR_j$ for trading rule *j* at time *t* is defined by:

$$SR_{j,t} = \frac{\overline{f}_j}{\widehat{\sigma}_j}, \ j = 1, \ldots, l,$$

where $\overline{f}_{j,t}$ is the mean excess return and $\widehat{\sigma}_{j,t}$ the estimated standard deviation of the mean excess return. Another important feature of the Sharpe ratio metric is its direct link with the actual *t*-statistic of the empirical distribution of a rule's returns (Harvey and Liu, 2015)[3]. Thus, such a property makes the Sharpe ratio the most appropriate criterion for our proposed multiple hypothesis testing framework. In other words, the test statistic (i.e. $\varphi_j$) is the Sharpe ratio[4].

Through the specifications described, the performance of each rule is calculated and tested for significant positive difference compared to a benchmark. Following Sullivan *et al.* (1999) and Bajgrowicz and Scaillet (2012), our benchmark is the risk-free rate that corresponds to abstaining from the market when no profitable opportunity is expected. Alternatively, the benchmark can be defined as the buy and hold strategy on the MSCI World index or further to a combination of bonds and stock indices[5].

*The returns of the trading rules should be adjusted for exposure to known risk factors,*

*estimated from asset pricing models commonly used in the literature and in practice.*

As they do not exist common (or market specific) risk factors for all market indices under study that are recognized and accepted from the related literature, the returns of our technical rules cannot be adjusted for exposure based on them. The evaluation of the performance of our portfolios match that of portfolio managers and traders and of the related literature in technical analysis on multiple market indices (see amongst others, Sullivan *et al.,* 1999 and Hsu *et al.*, 2010)

## 3. Data Snooping Bias and Existing Methods

Forecasting financial series and assessing the profitability of a series of competing models can be considered to be the oldest and most popular research exercise in finance. Financial economists and practitioners deal with hundreds of competing models when it comes to the predictability of different assets or the true profitability of trading strategies. In order to distinguish the genuine and significant ones from the insignificant, economists and practitioners have to employ MHT frameworks instead of the classic statistical inference, which can lead to biased estimates. It is very likely that Type I error (i.e. probability of making false rejections) will emerge when multiple hypotheses are tested. Classic statistical inference fails to capture the exact number of false rejections. MHT specifications attempt to control the number of Type I errors in a large universe of jointly tested hypotheses. The most well-established MHT approaches are the Family-Wise Error Rate (FWER), the False Discovery Proportion (FDP) and the False Discovery Rate (FDR).

---

[3] The t-statistic of a given sample of historical returns $(r_1, r_2, \ldots r_t)$, testing the null hypothesis that the average excess return is zero, is usually defined as $t = \frac{\hat{\mu}}{\hat{\sigma}/\sqrt{T}}$, while the corresponding Sharpe ratio is given by the formula $SR = \frac{\hat{\mu}}{\hat{\sigma}}$.

[4] We observe similar trends in our results with the Sortino ratio, the manipulation ratio and the annualized return as performance metrics. Sastry (2018) finds similar power for all four performance metrics.

[5] The choice of a relevant benchmark is central to the hypothesis testing and the set of discoveries. Although different possible specifications could be considered, the scope of this study is to propose the new MHT procedure in the most common and verified setting.

FWER is defined as the probability of having at least one Type I error. In other words, it measures the probability of having at least one false discovery. A testing method is said to control the FWER at a significance level $α$ if FWER$≤α$. Naturally, when a researcher tests a large number of hypotheses, it is highly likely that at least one Type I error will occur. There are several approaches to controlling the FWER. The most naïve method of doing so is the Bonferroni correction. In this approach, the researcher has simply to divide the number of tests from the significance level $α$. Then each test is run with a significance level $α/l$ (where $l$ is the total number of tests). The larger the number of tests, the smaller the common critical $p$-value. The Bonferroni correction is characterized by its simplicity but has been criticized for loss of power and a high probability of Type II errors (Benjamini and Hochberg, 1995). A less strict approach to controlling the FWER is Holm's stepwise method (1979). Here, the null hypothesis is rejected when $p_i \leq α/(l - i + 1)$ for $i = 1 \ldots l$. The criterion becomes less and less strict for large $p$-values and thus Holm's method rejects more hypotheses than the Bonferroni correction. However, it is worth noting that both methods ignore the dependence structure of the individual $p$-values, which makes them overly conservative.

Motivated by this fact, White (2000) introduced the Bootstrap Reality Check (BRC). In this approach, the FWER is asymptotically controlled by estimating the sampling distribution of the largest test statistic and taking into account the dependence structure of the individual test statistics. BRC applies bootstrapping to get less conservative critical values than the aforementioned approaches. The main disadvantage of BRC is that it only checks if the model or strategy that appears best beats the benchmark. It also has low power when strong underperformers exist in the hypothesis testing pool and the relevant p-values are still conservative (Romano, Shaikh and Wolf, 2008). To overcome this problem, Hansen (2005) introduced the superior predictive ability (SPA) test, which uses studentized test statistics and attributes lower weights to those corresponding to poor performers. However, this approach also only focuses on the performance of the best strategy. Moving one step forward, Romano and Wolf (2005) introduced the StepM test (RW) in an attempt to statistically validate as many outperforming strategies as possible. The RW test improves upon the BRC just as the stepwise Holm method improves the single-step Bonferroni approach. The RW test initially identifies the most robust strategies through a stepdown approach, until a false selection is observed. The first step of the RW is the same as in the BRC test. In the next step, the remaining strategies are again evaluated over a new critical value (based on bootstrap) and these iterations continue until no further strategies are rejected. Likewise, Hsu, Hsu and Kuan (2010) introduced a stepwise version of Hansen's SPA test (2005) to control any negative effect of underperforming strategies on the power of the test. Given that RW and also the method of Hsu *et al.* (2010) are based on bootstrap estimates, it is safe to assume that it is less conservative, in spite of the correlation structure of the *p*-values and still asymptotically control FWER (as BRC does). However, it still remains a strict approach, since the procedure terminates once a false rejection is identified. To solve this issue Romano, Shaikh and Wolf (2008) relaxed the strict FWER criterion by introducing the $k$-FWER method (similar to a $k$-StepM approach). The innovation of this method is that it allows for a $k$ number of false rejections before it stops. As long as the false selections are less than $k$ the procedure continues in subsequent steps similar to RW. This makes the outcome less conservative, but the results are quite sensitive to the selection of $k$.

The FDR is based on the idea of allowing for a specific number of false negatives when a practitioner observes quite a large number of rejections, which increases the power of the test while relaxing the testing framework. Introduced by Benjamini and Hochberg (1995) as a more tolerant error metric, the FDR measures the proportion of false discoveries among true rejections of the null hypothesis. Specifically, they suggest that if *F* and *R* are the number of the total Type I errors (false discoveries) and the rejected null hypotheses (total discoveries) respectively, then the FDR is estimated as $FDR = E(F/R)$. Benjamini and Hochberg (1995) conclude that if all tested null hypotheses are true, then FDR is equivalent to FWER. However, if the number of true discoveries is lower than the total null hypotheses tested, then FDR is smaller than FWER. In addition, the FDR corresponds to the expected FDP, or in other words, it controls the FDR at level $\gamma$ (i.e. $FDR = E(FDP) \leq \gamma$)[6]. Over the years many studies have tried to control the FDR measure, some with more incremental and others with more comprehensive approaches. Nevertheless, the fundamental idea of identifying as many true rejections as possible without including too many false ones remains the same (Benjamini and Yekutiely, 2001; Storey, 2003; Storey and Tibshirani, 2003; Storey et al., 2004; Liang and Nettleton, 2012; Liang, 2016). In financial applications, Barras et al. (2010) introduced for the first time an FDR approach similar to that of Storey (2003) which focuses on measuring the proportion of false discoveries among mutual funds generating alphas, while trying to identify those displaying significant positive performance.

---

[6] $\gamma$ is user defined and should not be confused with $\alpha$.

The relevant literature highlights the superiority of the FDR process (see among others, Harvey and Liu, 2015; Bajgrowicz and Scaillet, 2012; Liang, 2016). The advantage of this method lies in the fact that by tolerating a certain, usually small amount of Type I errors, the FDR improves the power to detect more significant discoveries, compared to its stricter competitor, the FWER, which guards against a single erroneous selection and so leads to missed findings. This implies that lower critical values are used and consequently a larger number of significant strategies is selected. This is particularly important in finance and trading applications, as in practice investors prefer several alternative strategies, rather than basing their entire strategy on a single trading tool. Additionally, the FDR test takes into account all the outperforming rules in the population and it does not terminate when a single rule, even the best, yields a lucky performance. Therefore, conceptually the FDR application is more suitable when analyzing large data sets and aiming to make reliable statements about the realized average FDP across the various data sets, as in our case (Benjamini, 2010). Concerning the FDP, controlling it can lead to more conservative estimates compared to the respective FDR ones (Genovese and Wasserman, 2006). Sun et al. (2015) suggest that in conditions of strong dependence FDP can be highly volatile. Additionally, as Fan and Han (2017) find that FDP estimates can be relatively small for large data sets, the FDR approach is more suitable for our big data analysis.

## 4. Methodology

### 4.1 Overview of the FDR Procedure

As mentioned previously, the FDR is defined as the proportion of false discoveries among the rejected null hypotheses. FDR is an expectation and thus its control does not require an additional specification at a probabilistic level (as for example, the FWER does). Methods to control the FDR have been suggested by Benjamini and Hochberg (1995), Benjamini and Yekutieli (2001) and Storey (2002). Benjamini and Hochberg's method (1995) assumes that the *p*-values are mutually independent, something that is not the case in our study. On the other hand, the Benjamini and Yekutieli (2001) approach assumes that *p*-values have a more arbitrary dependence structure, but it is less powerful. Storey (2002) improves its power with an approach based on the assumption that, for a two-tailed test, the true null *p*-values are uniformly distributed over the interval [0,1], whereas the *p*-values of alternative models lie close to zero. His approach utilizes information from the centre of the distribution of *t*-statistics (i.e. $\varphi_j$), which is mainly represented by non-outperforming rules. A key point regarding this direction is the precise estimation of the proportion of rules satisfying the null hypothesis, $\varphi_j = 0$, (i.e. $\pi_0$) in the entire population. A conservative estimator of the $\pi_0$ parameter is given by:

$$\widehat{\pi_0}(\lambda) = \frac{\#\{p_j > \lambda; \ j = 1, \ldots, l\}}{l(1 - \lambda)}$$

where $\lambda \in [0,1)$ is a tuning parameter indicating above which specific level the null *p*-values exist. The required inputs for the FDR approach are mainly the (two-sided) corresponding *p*-values of the performance metrics ($\varphi_j$) of each individual rule associated with null hypothesis of non-abnormal performance ($H_{0j}: \varphi_j = 0$) against the alternative of abnormal performance ($H_{Aj}: \varphi_j > 0$ or $H_{Aj}: \varphi_j < 0$). Furthermore, there is no need for a priori knowledge of the *p*-values distribution. The stationary bootstrap resampling technique of Politis and Romano (1994) is applied to obtain the individual *p*-values. It is applicable in cases where the time series are weakly dependent (which is the case in technical rules performance).

We incorporate into our approach the FDR$^{+/-}$ method of Barras, Scaillet and Wermers (2010), which provides separate estimates of the percentages of false discoveries among outperforming and underperforming rules compared to the benchmark. However, since we are interested in identifying only the positive outperformers in our application, we focus on the case where $\varphi_j > 0$. In the context of the performance of technical trading rules, the FDR$^+$ is described as the expected value of the proportion of erroneous selections, $F^+$, over the significant and positive rules, $R^+$, (i.e. $\frac{F^+}{R^+}$). The number of $F^+$ represents the rules, whose *p*-values falsely reject the true null (i.e. $H_{0j}: \varphi_j = 0$) in favour of the alternative and exist among $R^+$. On the other hand, $R^+$ portrays the number of rules rejecting the $H_{0j}$, in a two-tailed test, and their performance metric $\varphi_j$ is positive. The estimate of FDR$^+$ is given by $\widehat{FDR}^+ = \widehat{F}^+ / \widehat{R}^+$ where $\widehat{F}^+$ and $\widehat{R}^+$ are the estimators of $F^+$ and $R^+$, respectively[7]. For example, an FDR$^+$ of 100% conveys that no trading strategy genuinely outperforms the benchmark, while any existing performance can be purely attributed to chance. In general, the FDR produces a sensible trade-off between true positives and false selections, while it is less conservative than the FWER measure in terms of power. Because of this less conservative nature, the FDR method has the advantage of selecting outperforming rules even if the best rule is not significant in terms of performance.

We can estimate the frequency of false discoveries or the number of lucky rules, $F^+$, in the right tail of the distribution of performance metrics, $\varphi_j$ at a given significance level $\gamma$ as:

$$\widehat{F}^+ = \pi_0 * l * \gamma/2$$

where $\pi_0$ is the proportion of rules satisfying the null hypothesis, $\varphi_j = 0$, in the entire population, $l$ is the number of the entire population and $\gamma/2$ is the probability of a positive non-genuine rule exhibiting luck due to symmetry conditions.

## 4.2 Issues regarding False Discovery Rate Existing Methods

Before describing our proposed approach, we provide a discussion of possible issues arising in both the multiple hypothesis set-up and the control of FDR in this section. Benjamini and Hochberg's original idea about FDR (1995) assumes that the multiple hypotheses tested are independent of each other. However, a considerable number of trading rules in our sample are just small variations of themselves. For example, moving averages are highly correlated since we consider only slightly different parameters during their construction. Efforts have been made to provide evidence for "weak dependence" conditions of the test statistics under which the FDR approach holds (Benjamini and Yekutiely, 2001; Storey, 2003; Storey and Tibshirani, 2003; Storey et al., 2004; Farconemi, 2007; Wu, 2008). Most of them show that this happens when the number of tests increases to infinity with dependence effects diminishing to zero due to asymptotics. Likewise, in our empirical investigation the technical trading rules display dependence within specific classes (i.e. moving averages), while each class is independent of each other (i.e. different families of rules). For this purpose, we also need to confirm whether a weak dependence condition holds for our discrete FDR framework before we move forward to construct it as well as test the behaviour of our FDR approach under cross-sectional dependences in a Monte Carlo simulation.

---

[7] Similarly, the $\widehat{F}^-$ and $\widehat{R}^-$ correspond to the estimators of the proportion of false discoveries ($F^-$) and alternative rules ($R^-$) generating negative performance (i.e., $\varphi_j < 0$).

Another important issue arises when we have to perform hypothesis testing in a large number of *t*-statistics, usually numbering in the thousands, while the number of observations is relatively small, as it is in our case. Specifically, we utilize more than 21,000 technical trading rules over an IS horizon of two years (i.e. 504 observations). Bootstrapping procedures, as mentioned above, are commonly used in these cases to calculate the corresponding *p*-values and so to perform hypothesis testing, due to the fact that they require few distributional assumptions and are robust to outliers. Performing a resampling procedure on each trading rule though, generates *p*-values which are discrete rather than continuous because of the finite number of bootstraps employed. This leads to the detection of large-scale homogeneous discrete *p*-values, sharing the same support points. Previous studies either controlling FDR or FWER overcome this issue by assuming that the true null *p*-values are continuous and follow a uniform distribution as described above (Storey, 2002; Storey *et al.,* 2004; Romano and Wolf, 2005; Romano and Wolf, 2007; Barras *et al.,* 2010; Brajgowicz and Scaillet, 2012). However, the true null discrete *p*-values tend to be stochastically larger than uniformly distributed and the direct application of existing methods to them can cause several misspecifications (Pounds and Cheng, 2006).

Moreover, in a two-sided test and for continuous true null *p*-values uniformly distributed, it holds $\Pr(p-\text{value} \leq \gamma) = \gamma$ for all $\gamma \in [0,1]$. On the other hand, for the discrete ones we observe only a certain number of support points for the $p-values$ (i.e. $V = \{\gamma_1, \dots, \gamma_v, \gamma_{v+1}\}$ with $0 < \gamma_1 < \dots < \gamma_v < \gamma_{v+1} \equiv 1$) with potentially many ties, which satisfy a *discrete uniform condition* such that $\Pr(p-\text{value} \leq \gamma) = \gamma$, for $\gamma \in (0,1]$ and only for $\gamma \in V$. Using bootstrapping techniques to compute the associated *p*-values for each rule, we end up with *p*-values satisfying a discrete condition with common support points. To illustrate further, every *p*-value is usually calculated by comparing the value of each performance metric with the value of its corresponding quantiles of bootstrapped metrics (Sullivan *et al.,* 1999). This means that large values of observed test statistics provide evidence against the null and the corresponding *p*-values are given as $p_j = \frac{1}{B}\sum_{i=1}^{B}(\varphi_{jb} \geq \varphi_j)$, where *B* is the number of bootstrap replications, while $\varphi_{jb}$ is the test statistic calculated for the *b*th bootstrap for the *j*th rule and $\varphi_j$ is the realized test statistic. *P*-values computed this way are attached with support points in the form: $V =$

$\left\{\frac{1}{B}, \frac{2}{B}, \ldots, \frac{B-1}{B}, 1\right\}$, which also verify a discrete nature. Thus, providing an FDR framework, which takes into account larger discrete *p*-values (opposite to smaller, continuous ones) might help improve existing methods further.

Another issue appears in the calculation of $\lambda$ parameter and so in the estimation process of $\pi_0$, which is the key estimator for controlling FDR. Generally, poor selection of $\lambda$ can cause unnecessary conservativeness in $\widehat{\pi_0}$ and $\widehat{FDR}$. For example, not all values of $\lambda \in [0,1)$ generate ideal $\widehat{\pi_0}$ estimates under a discrete set-up, contrary to what usually happens in a continuous framework. Imagine a candidate set of $\lambda$, $L = \{\lambda_0, \lambda_1, \ldots, \lambda_v\}$ with $\lambda_0 \equiv 0$. We can show that, if we arbitrarily select $\lambda$ from $L$ for some fixed value $q \in \{0, \ldots, v\}$ based on a support point, then $\widehat{\pi_0}(\lambda)$ is a conservative estimator of $\pi_0$. On the other hand, if $\lambda$ does not belong to this set but lies in between two support points (e.g. $\lambda_i < \lambda < \lambda_{i+1}$)), then choosing $\widehat{\pi_0}(\lambda) > \widehat{\pi_0}(\lambda_i)$, can lead to an extra and unnecessary conservativeness in the estimation of proportion of rules with no abnormal performance (Harvey and Liu, 2018).

In terms of choosing $\lambda$, a small value can lead to estimators with a large positive bias, while a high value of $\lambda$ leaves only a small number of *p*-values on its right-hand side to estimate $\pi_0$, yielding an increase in the variance of estimators. Thus, we should always achieve a good trade-off between the two when choosing $\lambda$. Previous literature follows a common approach to choosing $\lambda$ under a continuous set-up, visually examining the histogram of all *p*-values and setting the $\lambda$ parameter equal to the support point above which the number of occurrences of *p*-values becomes fairly flat or selecting an arbitrary value (most commonly λ=0.6 based on the suggestion of Brajgowicz and Scaillet, 2012). The rationale is based on the assumption that bootstrapped *p*-values share equally spaced support points and each support contains a uniform number of true null *p*-values.

As mentioned above, previous studies conclude that the conservativeness of $\widehat{\pi_0}$ holds for almost every fixed $\lambda$ approximately satisfying this condition. However, we find in our extensive Monte Carlo simulation that $\widehat{\pi_0}(\lambda)$ is sensitive to some extent to the choice of $\lambda$, a finding that is contrary to previous evidence produced by Brajgowicz and Scaillet (2012) and in agreement with the simulations of Harvey and Liu (2018)[8]. Furthermore, such a technique might also involve an extra bias depending on how the researcher interprets a specific level of a histogram's flatness. For the above reasons, we concentrate in the next section on selecting $\lambda$ dynamically, as a fixed quantile of *p*-values, based on the data characteristics, to minimize any conservativeness that may be undesired by the estimators.

---

[8] Brajgowicz and Scaillet (2012) explored three values, λ=0.4, 0.6 and 0.8. The authors found the value 0.6 to be optimal and they note that their results are not sensitive on the choice of λ and their exercise on the DJIA index. Harvey and Liu (2018) with a similar application that explores the same set of values for λ (0.4, 0.6 and 0.8), produce results that are sensitive to the choice of λ on Standard and Poor's Capital IQ database. On their database the optimal value of λ is 0.8. In our application where we study multiple series, we also note that the results are sensitive to the choice of λ and that its optimal value in our approach differs between the series and periods.

Finally, MHT frameworks are computationally demanding most of the time, since they involve bootstrapping procedures. Moreover, the FDR approach requires setting the tuning parameter by taking into account the graphical representation of *p*-values, which considerably increases the computation time needed for controlling FDR. Our proposed dynamic approach is computationally more efficient in terms of both time and selecting the outperforming rules based on an algorithmic set-up, which can also help practitioners make better decisions in portfolio construction and OOS estimation.

### 4.3 DFDR$^{+/-}$

We now present our novel specification DFDR$^{+/-}$ accounting for homogeneous discrete *p*-values, while providing different estimates of the proportion of false selections for outperforming and underperforming rules. However, since our aim is to identify the significant outperforming rules only, we will focus on the estimation of DFDR$^+$ for the rest of this paper. In general, our study is the first in the field of finance to propose an adaptive FDR approach employing a dynamic parameterization, while considering discrete *p*-values as a tool for controlling data snooping.

Our approach concentrates on large-scale homogeneous discrete *p*-values. Following Kulinskaya and Lewin (2009), we assume that using bootstrapping procedures as described in Section 4.2, we acquire discrete *p*-values which satisfy a uniform condition, while sharing the same discrete support *V*. Furthermore, we need to consider as $N = \{n_1, \ldots, n_{v+1}\}$ the number of occurrences of each element in *V*, i.e. $n_i = \#\{p_j = \gamma_i\}$ for $i = 1, \ldots, s + 1$ in order to express the empirical distribution of the computed *p*-values. Thus, the empirical distribution of homogeneous discrete *p*-values with common support points is totally explained by $(V, N)$. Coming to the FDR approach calculation, the $F^+$, $R^+$ and FDR$^+$ also represent step functions with possible change points at the support points. Then it is sufficient simply to acquire their values at the specific support points to control the FDR[9]. Given that, the distribution function of the null discrete *p*-values on every support point is identical with that of continuous *p*-values, which is the key evidence in developing a parallel method for discrete *p*-values based on similar dynamic set-ups for continuous *p*-values.

---

[9] Suppose that $\gamma$ is a time-running parameter from zero to one, then the continuous time processes $F^+$, $R^+$ and FDR$^+$ relax to discrete stochastic process on the support points.

We now explain our new approach to improving the FDR$^{+/-}$ methodology to accommodate for discrete *p*-values, while dynamically selecting λ under a stopping time rule, similar to viewing the time running forward. We define this stopping time condition as the point which holds $E[\widehat{\pi_0}(\gamma_q)] \geq \pi_0$, while *q* is the exact stopping time with respect to $n_i$ (for $i = 0, ..., s$), which is the history of *p*-values up to $\gamma_q$, when $q = i$. We also determine the whole procedure up to the stopping time *q*, as $\{0 \equiv n_0, ..., n_i\}$. Then we just set λ equal to $\gamma_q$. We check every support point instead of checking every single *p*-value for the stopping condition. If *q* is an appropriate stopping time, it must also hold $E[\widehat{FDR}(\gamma_q)] \geq FDR$, where $\widehat{FDR}$ is the estimation of the actual FDR provided by our method. The rationale for this approach is related to the idea of discovering the smallest support point, in which the number of appearances of *p*-values, $n_i$, to each right-hand side is almost equal. However, the stopping time condition is very general since we can construct numerous stopping time rules fulfilling the above criteria, while the actual right-hand side counts are unobservable, setting hurdles in the computation of the stopping time approach. It is already known that employing a right boundary procedure, such as the one introduced by Liang and Nettleton (2012) for continuous *p*-values solves this issue by only taking into account the average of the remaining counts. In general, the right-boundary specification guarantees conservative estimators for $\pi_0$ and FDR depending on a grid of candidate points for λ in line with data characteristics and a stopping time condition, at least for a continuous framework (Liang and Nettleton, 2012). We adopt the same procedure for discrete *p*-values. In addition to this, the right boundary procedure performs effectively for both independent and weakly dependent *p*-values, as observed in our case (see Liang and Nettleton, 2012; Liang, 2016). Liang and Nettleton (2012) and Liang (2016) provide evidence of computing an FDR estimator using the right boundary procedure, while certain limits exist. Their results clearly satisfy a special case of the weak dependence condition of Storey *et al.* (2004).

The aim of the right boundary procedure is to find the first λ, at which the values of $\widehat{\pi_0}(\lambda)$ stop decreasing, satisfying in that way the stopping time condition. For this reason, we consider a candidate set for λ, $\Lambda = \{\lambda_1, ..., \lambda_n\}$, in which we place its components in ascending order, $0 \equiv \lambda_0 < \lambda_1 < ... < \lambda_n < \lambda_{n+1} \equiv 1$ (and $\lambda \subseteq \Lambda$). Then we select the best λ, as the minimum $\lambda_q$, which fulfils the condition $\widehat{\pi_0}(\lambda_i) \geq \widehat{\pi_0}(\lambda_{i-1})$, (i.e. $q = \min\{1 \leq i \leq n-1 : \widehat{\pi_0}(\lambda_i) \geq \widehat{\pi_0}(\lambda_{i-1})\}$). Specifically, we use the set $\Lambda$ to separate the interval between zero and one, $(0,1]$, into $n+1$ bins with the $i$-th bin being $(\lambda_{i-1}, \lambda_i]$ for $i \in \{1, \ldots, n+1\}$ and $w_i = \#\{p_j \in (\lambda_{i-1}, \lambda_i]\}$ being the number of $p$-values in the $i$-th bin. Assuming equal intervals between $\lambda$s, this approach practically chooses the right boundary of the first bin whose number of $p$-values is no larger than the average of the corresponding number to its right. In this way, we achieve the stopping condition when the downward trend of the number of $p$-values in each bin is neutralized, as we move forward, to a level where the random variants of rest $p$-values are fairly equal. Finally, acquiring the optimal $\lambda$ in this way, we can easily calculate a conservative estimator for $\pi_0$ based on Storey's formula (2002) as has already been mentioned in previous sections.

The rest of the steps for the selection of outperforming rules remain similar to those of Barras *et al.* (2010) in the FDR specification. In terms of bootstrapping, we generate 1,000 sequence replications, and we retain the same bootstrap draw of the time series sample period for each trading rule's returns. In this way, we actually bootstrap the cross-section of trading rules returns through time in order to preserve the cross-sectional dependencies (Kosowski *et al.*, 2006; Fama and French, 2010; Yan and Zheng, 2016). The application of stationary bootstrap also allows us to preserve the autocorrelations in returns structures. We then use the "point estimates" procedure of Storey *et al.* (2004) on generated $p$-values, under weak dependence to select the outperforming rules, while setting a target for false discoveries. We can also extrapolate the proportion of trading rules displaying non-zero performance as $\pi_A = 1 - \pi_0$ in the entire universe of technical trading rules by using the FDR approach. This may be useful for an investor who wants to divide $\pi_A$ into the proportions of positive, $\pi_A^+$, and negative, $\pi_A^-$, rules in the population. The former includes both alternative rules generating positive performance and rejecting the null ($p-value < \gamma$) as well as those with positive performance but not rejecting the null ($p-value > \gamma$). The latter include those relevant for rules showing negative performance. We describe in Appendix A the precise steps for achieving this, the estimation of $\lambda$ and $\widehat{\pi_0}$, as well as the computation of $\pi_A^+$ and $\pi_A^-$ in . In our Monte Carlo simulation, also presented in Appendix A, we provide evidence that our discrete right boundary FDR procedure achieves a good trade-off between the bias and variance in various weakly dependent settings. We also compare the performance of the proposed procedure relative to the established FDR procedure of Storey *et al.* (2004) as well as the StepM test (RW) of Romano and Wolf (2005).

### 4.4 DFDR portfolio construction

We construct a portfolio of technical trading rules by setting the $\widehat{DFDR}^+$ target (the estimated DFDR for outperforming rules only) to 10%, which achieves a good trade-off between wrongly chosen rules and truly outperforming ones. In particular, our Monte Carlo simulations reveal stability when $\widehat{DFDR}^+$ levels range from 5% to 30%. For our 10%-DFDR$^+$ portfolio, only 10% of the rules selected do not have genuine profitability among the outperforming rules, while 90% possess significant predictability. Moreover, we use the forecast averaging technique and allocate equal weight to the signals pooled from the chosen rules at each time-step in order to construct and calculate the portfolio returns. Since each trading rule might generate a long, short or neutral signal at a single time-step, we invest an equal proportion of our wealth in the signals and their corresponding returns generated by each individual rule, similar to calculating their equally-weighted cross-sectional mean.

Following previous studies (see among others, Brock *et al.*, 1992; Bajgrowicz and Scaillet, 2012), a trading position is opened when a long or short signal is produced and liquidated when the signal is either reversed or neutralized. Should a neutral position be raised, the proportional wealth is assumed to be invested in the risk-free asset or the saving account. The gross daily return is calculated by the change in the closing value of the underlying index. A one-way transaction cost is deducted from the gross return when a position is terminated. The excess return is then estimated, to compare the profitability of the trading rules with the risk-free rate.

### 5. IS Performance

This section provides an ex post analysis for the technical rules. In Section 5.1 the number of strategies identified as genuinely profitable for the two-year look-back period, their relevant profitability and a disaggregated analysis of the types of trading rules that are profitable is presented. Section 5.2 reports their corresponding break-even transaction cost analysis. The relevant results based on one-year IS are presented in Appendix B.1.

### 5.1 Identification, IS Profitability and Disaggregation Analysis

Table 2 presents the percentage and standard deviations of the survivor rules identified by our 10%-DFDR$^{+/-}$ portfolios. The look-back period is set to two years and the portfolios are readjusted on a monthly basis.

[Table 2 here]

We note that the percentage of identified rules varies over the years and from market to market. The higher number of identified rules are found in the UK, Russia and the Frontier markets indices. There is no obvious trend in the percentage of identified rules over the years. The peaks are in 2006, 2009 and 2010 while the lowest is in 2012. It is interesting to note that our procedure selects profitable rules for all indices and all years. The relevant trading performance of the portfolios generated during the look-back period is presented in Table 3.

[Table 3 here]

There is a significant profitability pattern after transaction costs for all indices. Emerging markets present an increased profitability compared to their counterparts in terms of annualized return. There is no obvious trend in the profitability of technical analysis. There is a peak in terms of annualized returns for the years 2009 and 2010 and consistent stable Sharpe ratios for all years. There is also no connection between the percentage of identified rules and the trading performance of the generated portfolios.

Table 4 reports a disaggregated analysis of the classes of trading rules that are profitable (i.e., RSIs, filter rules, moving averages, support and resistance and channel breakouts) in every single market based on the findings of Table 2. In Table 4 we present the average percentages of the families of rules that are profitable across all years (i.e. 2006-2015) when the lookback period is set to two years and the portfolios are readjusted on a monthly basis[10].

[Table 4 here]

It is obvious that moving averages dominate their counterparts across all markets. The second largest family of genuine profitable rules is that of support and resistance. We note that contrarian rules (RSI) present considerably lower percentages for all markets under study (with the notably exception of Jordan and UK). It seems that all markets under study are characterized by momentum and strong trends. Our results extend the findings of Chan *et al.* (1997), Jegadeesh (1990) and Jegadeesh and Titman (1993, 2001) who find profitability on momentum trading strategies on US and European stocks.

## 5.2 Break-even Transaction Costs

In this section, we perform a break-even analysis of the excess profitability of technical trading rules over the look-back period. Following relevant studies in the field (see among others, Bessembinder and Chan, 1998; Bajgrowicz and Scaillet, 2012; Hsu *et al.,* 2016), we adopt as the break-even cost the level of one-way transaction cost, which makes the excess profitability (i.e. mean return) of the best-performing technical trading rule diminish to zero. The more the break-even costs surpass the actual costs, the more robust a rule's excess profitability is deemed to be.

---

[10] A more detailed disaggregated analysis of the classes of trading rules that are profitable for every single year and market is available on Appendix B.2.

Figure 1 displays the size of average break-even transaction costs for the best-performing technical trading rule under the Sharpe ratio metric and for each index separately. We select the best-performing rule for every month based on the two-year look-back period that acts as the IS[11]. Then the procedure is repeated for each of the 12 months in every year, while rebalancing is performed. The average break-even transaction cost per year is estimated by dividing the sum of the best rule's monthly break-even transaction costs by twelve. The same procedure for the overall 10-year period is applied.

[Figure 1 here]

The major trend revealed by the figure is that frontier markets achieve the highest break-even transaction costs, followed by the emerging and advanced markets respectively, at least for the first four years. In particular, Morocco (up to 38%), Russia (up to 43%) and Brazil (up to 41%) dominate in terms of excess profitability robustness over that period, while on the other hand an advanced market such as Japan (up to 52%) reports the highest break-even costs over 2009 and 2011. For the rest of the study periods, there is a decay of break-even transaction costs, except from 2014, from which point the advanced markets recover compared to the rest. For instance, the break-even transaction costs of the corresponding advanced markets' index as well as the US span from 14% to 20%. Of course, there are still some emerging (i.e. Russia and Brazil) and frontier (i.e. Estonia and Morocco) markets, which score similar or even higher break-even costs.

In general, we observe a downward trend of break-even costs and so excess profitability over the years, costs reaching their lowest levels in recent years and especially in 2015. However, this trend is not always stable. Specifically, most of the countries exhibit high break-even costs in 2006, facing a small decay in 2007, while they recover to higher levels from 2008 to 2010, years during which they also reach a characteristic peak. During the following years there is a considerable decay in their size, with a slight recovery only in 2014, which does not clearly remain in 2015. So, apart from the years in which break-even costs report their highest values (i.e. 2008-2010), there is a somewhat consistent performance of technical trading rules especially in recent periods.

---
[11] The corresponding break-even transaction costs using one year as IS reported in Appendix B.3 reveal lower break-even transaction costs in general. However, the major trends found in Figure 1 remain.

## 6. OOS Analysis

This section provides a comprehensive analysis of the portfolios constructed on the surviving rules of the DFDR$^+$ procedure with an ex-ante approach. Section 6.1 presents the excess profitability of the DFDR$^+$ portfolios over the OOS. Section 6.2 studies performance persistence by measuring the number of periods a DFDR$^+$ portfolio can generate a return above the risk-free rate. In Section 6.3, a novel cross-validation practice is provided that considers both IS and OOS and preserves the order of the underlying time series. Finally, Section 6.4 focuses on interpreting the excess returns of the portfolio over the different period by considering the level of financial stress in markets.

### 6.1 Profitability

Following previous studies in the field, we employ an OOS experiment based on the performance of significant technical trading rules IS. We create portfolios of outperforming trading rules for each index, based on their IS significance, and we evaluate them in OOS. We consider three OOS periods of one, three and six months, following a two-year look-back period[12]. For the one-month post-sample period for example, we select the significantly positive rules based on their performance over the previous two years and under the 10%-DFDR$^+$ approach as a portfolio construction tool (see Section 4.4). Then the portfolios' performance is evaluated in the following one-month period. We rebalance our DFDR$^{+/-}$ portfolio every month in a rolling-forward structure over a year and we repeat the same procedure for all the years in our sample (i.e. 2006-2015). In this way we dynamically build and evaluate our portfolios just as an active investor would in practice. We utilize the OOS periods of three and six months respectively, in a similar manner.

Table 5 reports the average excess annualized mean return and Sharpe ratio (in parenthesis)[13] for every index, using a two-year look-back period as IS and one month as OOS over the full sample of ten years after transaction costs. Both annualized mean excess returns and Sharpe ratios presented are calculated as the corresponding OOS averages of twelve portfolios of significantly profitable outperforming rules built for every index after rolling forward the IS by one month during a year.

---

[12] We also set the IS period covering one year, but the corresponding portfolios of significant rules perform slightly worse than those constructed under a two-year IS period. The relevant tables and discussion are provided in Appendix B.4.

[13] We observe similar trends (time-varying profitability) with the Sortino ratio and the manipulation ratio. These results are available upon request. The results presented in Table 5 are the averages of 12 consecutive experiments for each year. This procedure and the fact that our OOS is short (even in the 3- and 6-month cases presented in the Appendix) validates the discussion that follows, irrespective of the performance metric.

[Table 5 here]

OOS evidence shows that technical trading rules perform quite well for the majority of the markets considered during the earlier periods (i.e. 2006, 2007), with the performance of emerging markets being more profound. Focusing now on the following years and especially 2008 and 2009, almost all markets present extraordinary performances. 2008 and 2009 correspond to the global financial crisis period, during which most of the markets faced extreme downward trends and so severe losses. This environment seems beneficial for our technical trading rules' universe, as it consisted mainly of momentum rules exploiting such big trends. This also explains the negative performance for most trading practices in 2010, which was a turning point for most of the markets. Additionally, the limited profitability of our DFDR$^{+/-}$ portfolios continues to a certain extent for the following years, and then it recovers for the more recent periods. Profitability over recent years (i.e. 2013-2015) is evident mostly in some advanced and emerging markets. When it comes to the comparison of the overall performance among markets, frontier markets provide a more solid investment option over the full sample period, generating positive mean returns and Sharpe ratios for the majority of the years, with second and third best being emerging and advanced markets respectively. Our general finding though, is that the profitability of technical trading rules is variable across the years examined and technical analysis can exploit short-term market inefficiencies.

Tables 6 and 7 display the annualized average excess mean returns and Sharpe ratios (in parenthesis) after transaction costs for every index using the two-year look-back period as IS but this time three-month and six-month OOS periods are applied, respectively. Once again both reported measures have been computed as the OOS averages of the corresponding twelve monthly portfolios per annum for every index in a three- and six-month rolling-forward structure, respectively.

[Table 6 here]

[Table 7 here]

As in the case of using one month as the OOS period, there is considerable evidence of excess profitability of technical trading rules during the earlier years, but this is now mainly concentrated on the emerging markets and only on a few of them. The outperformance reaches its highest level once again in 2008, with more than half of the indices switching to negative returns in 2009. This might be a result of the longer term OOS periods examined and so less frequent adjustment to new trends. Once again, the performance of technical trading rules diminishes over the upcoming periods, but technical analysis is still able to predict quite well some of the advanced market indices in very recent years, in terms of mean excess returns and Sharpe ratios. This may indicate specific patterns in market trends during these years, which are exploited by technical trading rules. Frontier markets still seem a good investment opportunity, with the emerging ones to follow across the years, even though the OOS period has now increased.

The previous results highlight the effectiveness of technical analysis in exploiting short-term market inefficiencies. As the OOS is increased, the profitability of our rules decreases, which is justified from both applications setting the look-back period to one (see Appendix B.1) and two years respectively. Such a result happens naturally as market efficiency eliminates any profitability and/or the underlying trends dominating the stock markets change. These findings can also explain parts of the past literature in the field of technical analysis and their profitability. Previous researchers have tended to examine long periods (see among others, Sullivan *et al.,* 1999; Qi and Wu, 2006; Bajgrowicz and Scaillet, 2012; Hsu *et al.,* 2016). As our empirical exercise highlights, the profitability of specific technical rules is short-term and varies across periods. Rules that survive for long periods naturally do not exist due to market efficiency. The profitability of specific technical rules seems short-term and adaptiveness might be the key.

### 6.2 Performance Persistence

In the previous section, we focused on identifying profitable technical rules with DFDR$^{+/-}$ and examining the trading performance of the generated portfolios. As discussed, it seems that technical rules and our approach are able to exploit short-term market inefficiencies. However, market efficiency should diminish any profitability sooner or later. It will be interesting to see how fast this profitability decays and whether there are differences between different markets and periods. This element is overlooked in the related literature, where the empirical evaluation is static and limited to specific periods. In real-world trading environments, practitioners are adaptive and rebalance their portfolios on a frequent basis.

Table 8 presents the persistence of our generated portfolios for the two-year look-back period (IS) and for the OOS one-month case. We measure persistence as the consecutive OOS months for which our portfolios have a trading performance above the relevant risk-free rate.

[Table 8 here]

We note that in the vast majority of cases, traders need to rebalance portfolios on more than a monthly basis. These results are expected, considering the level of efficiency of financial markets and the mediocre trading performance of our trading rules as outlined in the previous section. However, it should be noted that there is a handful of cases where the measured robustness is higher than one month. In other words, there are cases where market efficiency was weak enough to allow profitability for static portfolios. Persistence is higher for 2007 and 2008, an observation that can be attributed to the turbulence created by the global financial crisis and its side-effects on market efficiency. Interestingly, the market with the higher persistence is the US, the largest and most liquid index under study. In Table 9, we repeat the same exercise for the two-year look-back period and OOS three-month case.

[Table 9 here]

Persistence is decreased for most cases and years. We note that in 2008 the average persistence of our portfolios is more than 1. Else, our portfolios retain their profitability after the first three months in OOS. Considering specific markets, we note that the US retains a persistence larger than the OOS on average, which is consistent with previous findings. This is surprising, as one might expect frontier and emerging markets to have stronger persistence than advanced markets. Table 10 presents the same exercise for the two-year look-back period and six-month OOS.

[Table 10 here]

From Table 10, we observe that persistence is further decreased. There are years (especially for frontier markets) where persistence is 0. This means that none of the generated portfolios in these years have a trading performance higher than the relevant risk-free rate on the first six months of the OOS. Similarly, with the previous two cases, we note a peak for the year 2008.

The results above highlight the importance of rebalancing in trading. We note that there are a few cases where the portfolios might have negative profitability in the first month of the OOS, but some of them bounce back in the following periods (see for example, persistence for 2008 in Tables 8 to 10). In these cases, adaptiveness seems not to always lead to increased profits, while patience is rewarded. However, the majority of the results highlight the importance of rebalancing portfolios. Emerging and frontier markets do not seem to offer a safe haven for static portfolios. The observed trends allow us to note that persistence is stronger at the peak of the global financial crisis of 2008. These results lead us to further explore whether financial stress levels affect the profitability of technical analysis (see Section 6.4). This exercise also can be seen to show how adaptive a trader needs to be and how persistent the profitability of technical rules is. This choice can also be seen as a trade-off between Type I and Type II errors on the modelling part (Harvey and Liu, 2015), which is further examined in the next section.

**6.3 Cross-validation**

Splitting the data into IS and OOS plays an important role in the OOS profitability of technical analysis (see Sections 6.1 and 6.2).When only a small IS part out of the full sample data is retained for testing, the trading strategies that will be employed OOS, then it is highly possible for true discoveries to be missed (i.e. false negatives) (Harvey and Liu, 2015). Also, we are interested in identifying technical trading rules that are significant not only in IS but also in OOS. To validate that technical analysis has value in practice, we need to check whether technical rules exist that are profitable in the OOS and significant in both the IS and OOS.

Hence in this section we re-assess the robustness of our OOS performance in an innovative way based on a cross-validation experiment. In particular, we explore a method proposed by Harvey and Liu (2015), which involves a combination of full sample and the IS-OOS evidence in order to search for the intersection of survivors. Assuming that our IS period (i.e. two years) is not too short compared to the corresponding OOS periods (i.e. one, three and six months, respectively) we retain our IS-OOS test results computed previously. These involve the genuine technical trading rules which survived in OOS in terms of profitability (positive returns). At the same time, we employ the $DFDR^{+/-}$ method to select the significantly positive rules for a full sample horizon with a more lenient target rate (i.e. 20%). We consider three full sample periods corresponding to each of the three different OOS periods examined plus the IS period, while we utilize the aggregate data set each time (e.g. two years and one-month full sample in the first case). Finally, we merge the findings observed by the IS-OOS and the full sample simulations in order to identify the potential intersection of significant rules provided by the two approaches. Then we evaluate the performance of the rules belonging in the intersection. As far as we are aware, this is the first time such an approach has been adopted.

Tables 11-13 report the results of the cross-validation test as described above. In particular, the average OOS annualized mean excess returns of twelve cross-validated portfolios are constructed in a similar way to the previous sections after accounting for transaction costs. The average percentage of cross-validated rules in terms of the genuine significant rules, such as those identified in section 5.1, is also reported in parenthesis.

[Table 11 here]

Table 11 reveals the results of our cross-validation exercise for the one-month OOS case. Profitability is excessively high, especially during the global financial crisis (i.e. 2008-2009). For the remaining periods, emerging market indices demonstrate the highest performance and report very healthy returns with only a slight decay during the most recent years. The frontier markets indices exhibit similar patterns. For advanced markets indices, the performance of technical trading rules seems promising over most periods, while the returns yielded are quite low compared to those in advanced and frontier markets. Positive profitability was naturally expected, as cross-validated rules are a subset of the profitable OOS rules. The percentages of cross-validated rules vary considerably. For advanced indices these numbers span from 0.01% to 17.41% (i.e. both for the US) of the genuine IS rules, while indices for emerging markets range from 0.01% (i.e. China, Brazil) to 18.07% (i.e. China) across all periods. Moreover, the relevant percentages of cross-validating rules for frontier indices spread from 0.01% (i.e. Morocco, Jordan) to 18.14% (i.e. Morocco). On average, the percentage across all markets and periods is 2.47% of the related survivors presented in Table 2. The actual number of the rules is small (on average there are 30 rules), compared to the trading rules universe (i.e. 21,195). However, as noted before, the purpose of the exercise is to study whether technical analysis has any value in trading. The trading rules' universe considered is well diversified in terms of characteristics and parameters. In reality, practitioners take into account all available IS information and select a subset of rules based on this knowledge. For example, a trader may not consider any contrarian rules (i.e. RSIs) for OOS prediction in strong bearish/bullish periods such as the global financial crisis (i.e. 2008-2009). In other words, the question is whether there are rules that can survive the cross-validation exercise (if any) and not how many there are. The answer to that from our exercise is yes, since it finds that short-term market inefficiencies that can lead to profitability exist, and technical rules can exploit them. Table 12 presents the results of our exercise for the three-month OOS case.

[Table 12 here]

We observe the same characteristics in technical trading rules performance; however, the magnitude of generated returns is smaller compared to the ones obtained in the one-month OOS period. The percentage of cross-validated rules is at the same level as in the previous case. Table 13 presents the six-month OOS case.

[Table 13 here]

The evidence provided by performing the cross-validation experiment in a six-month OOS horizon reveals even lower returns compared to the relevant ones using one and three months across all indices and over all post-sample periods. This finding is consistent with the one in Section 6.1 describing the generic OOS results, due to the fact that the significant rules are exposed to longer post-sample periods with different trends.

The findings of this section reveal that genuine profitable technical rules exist, and technical analysis still has value in trading. Our DFDR$^{+/-}$ approach is capable of identifying subsets of these rules as demonstrated in Section 6.2. The percentages of cross-validated rules might seem small, but it is worth recalling that the scope of this section is to check whether truly profitable rules in both the IS and OOS do actually exist. These results can also be seen as the performance of an oracle trader that applies technical rules in our data set.

**6.4 Financial Stress**

In the previous sections, we noted a peak in the performance of technical rules for the years 2008 and 2009 that correspond to the recent global financial crisis. The performance of our portfolios deteriorates in the following years, but there are still cases where they present excess profitability after transaction costs even in advanced markets. Our results contradict the previous recent literature that finds no recent excess profitability after transaction costs for technical analysis (see among others, Hsu *et al.*, 2010; Bajgrowicz and Scaillet, 2012 and Taylor, 2014)[14]. These authors argue that the popularity of exchange-traded funds, algorithmic trading, market liquidity, derivatives or the effect of other macroeconomic factors have eliminated excess profitability in recent years. Although the effect of these factors cannot be rejected, our results guide us to explore the effect of financial stress on our portfolios.

In order to explore the effect of financial stress on our portfolios, we need first to measure it. The literature is rich in financial stress indices. The difference between them is based on the components that are used to construct them, their frequency and the market to which they are applied. In our study, we apply the Office of Financial Research (OFR) stress indices. They are constructed based on 33 market financial variables and cover the US, other advanced economies and emerging markets[15]. We match our findings for the US, Advanced and Emerging markets indices from Section 6.1 with the stress levels as reported by the related US, other advanced[16] and emerging markets OFR stress indices. Below we present the performance of our portfolios under high and low financial market stress[17].

---

[14] As discussed before, none of the previous literature presents such an extensive empirical application as the one presented here. Previous studies also applied more conservative and time-consuming MHT frameworks and limited their empirical applications to specific years or periods. For example, let us consider the case of the MSCI US index, the 2-year IS and 1-month OOS (see Table 5). If our application was limited only to 2014 or 2015, our interpretation would be anti-diametrical. The flexibility and adaptiveness of DFDR$^{+/-}$ along with recent developments in computational power allowed us to conduct an empirical analysis that unveils previously unknown patterns.

[Table 14 here]

We note that the trading performance of technical rules is considerably better when financial stress is high in emerging and other advanced markets. The annualized returns are five to six times higher in the high than the low period. This finding is consistent with the evidence found in Smith *et al*., (2018) where the authors demonstrate that technical analysis is relatively more useful in high-sentiment periods. During high-sentiment periods, financial markets exhibit stronger trends which can be captured by momentum based technical trading rules. The OFR stress index is based on series of indicators that capture financial market sentiment[18].

For US, we observe the opposite trend. The average profitability of technical rules over the 10-year period of study is positive in the US when stress levels are low and negative when stress levels are high. The US market is the most liquid and the biggest in terms of capitalization from the ones under study. It seems that under high financial stress periods, investors seek safe heaven to safer assets, switch to more complex algorithmic trading models or move to other markets. Our results can be explained by the level of sophistication of traders in US markets compared to their alternatives. In general, we note that financial stress seems to be an important factor in the performance of technical analysis, but its effect seems to depend on the relevant characteristics of the market.

---

[15] Other indices (such as the St. Louis Fed Financial Stress Index and the Kansas City Financial Stress Index) focus on a specific index while others stop before our sample (such as the International Monetary Fund stress index). Our criteria to select the index are to cover as many markets as possible from those under study and to ensure that all indices have been constructed with the same methodology.

[16] We note that the MSCI and the OFR indices do not correspond to each other perfectly. For example, the "other advanced" OFR index does not include the US. However, amongst the highest cited financial stress indices, OFR is the closest to our study.

[17] The OFR financial stress indices have a daily frequency and are volatile in certain periods. We examine only the 1-month OOS case as we are interested in measuring the effect of previous stress levels at the highest possible frequency.

[18] For a list of the OFR financial stress index sources see, https://www.financialresearch.gov/financial-stress-index/files/indicators/index.html.

## 7. Conclusions

Previous research on technical analysis' performance on equity markets is scarcely focused on certain features such as, study of a single, usually advanced, market index; a capped number and families of technical trading rules; no decomposition of their genuine profitable families; a non-realistic exercise of technical trading, which ignores transaction costs. In addition, other important questions, such as the length of the in-sample (IS)/out-of-sample (OOS) horizon for exercising technical analysis; the effect of financial stress on its performance; a detailed estimation of the persistence of technical analysis profitability; the importance of the frequency a portfolio of technical trading rules is rebalance, have rarely been answered or not answered at all. Hence, a comprehensive and up-to-date study of technical analysis on equity indices is more than urged from the academic community and the financial industry.

In this study we exercise a large universe of 21,195 technical rules on twelve MSCI advanced, emerging and frontier indices covering a period from 2004 to 2015. Towards this direction, we examine the IS and OOS profitability, its persistence, the role of financial stress and whether there are rules that are profitable and significant in both IS and OOS horizons. We also explore the role of the size of the IS and the OOS on technical analysis performance. In addition, we introduce a novel multiple hypothesis testing approach for controlling data snooping bias and so helping us to make accurate statistical inferences across a significantly large universe of technical trading rules. Our DFRD$^{+/-}$ adopts a large-scale homogeneous discrete *p*-values framework, while dynamically performing the estimation of its parameters. It is a fully adaptive computationally and efficient approach for data snooping testing that can assist academics and investors dealing with large data sets and a high number of competing models. A Monte Carlo simulation proves its accuracy and demonstrates its superiority in terms of power relative to a strong FDR benchmark.

The DFRD$^{+/-}$ approach identifies subsets of the technical rules that are genuine and profitable in the IS and the OOS. Among the families that outperform in all markets are mostly those of moving averages, channel breakouts and support and resistance, which belong to *momentum* classes of technical trading. Reversal strategies seem to have a weaker performance and significant presence only in specific markets (e.g., UK and Jordan). Additionally, the profitability and persistence of technical analysis vary in the OOS between the indices and the years, but we observe a peak for 2008 and 2009. In the following period, the profitability of technical analysis diminishes, only to recover again in the most recent period. Concerning the persistence of technical analysis, it is higher on the US index. We note that all our results are negatively affected when the IS is decreased to one year or the OOS is increased to three or six months. For the same US market, we also observe that technical analysis gains value when financial stress levels are low, while we get the opposite picture for the emerging and other advanced markets. Finally, a novel cross-validation exercise confirms that a small number of technical rules are genuine and profitable in both the IS and the IS-OOS periods.

Our results demonstrate that technical analysis still has value and can exploit short-term market inefficiencies even in advanced markets. Its profitability varies over the years and indices and seems to be a factor of several parameters. Our empirical exercise reveals two factors, the level of financial stress and the choice of the IS and OOS. Our result explains also to some extent the inconsistencies in the previous literature. Part of the previous research in the field tries to identify whether genuine technical rules exist for long periods (Brock *et al.*, 1992; Hsu *et al.*, 2010; Hsu *et al.* 2016). Our results demonstrate that the profitability of technical rules has short persistence. The fact that a technical rule is not profitable for extended periods does not mean that technical analysis has no value. Other research focuses only on individual indices and/or specific years for OOS testing (Brock *et al.*, 1992; Sullivan *et al.*, 1999; Bajgrowicz and Scaillet, 2012). As discussed before, technical analysis profitability varies over the years and according to indices, a fact revealed with the assistance of our novel $DFRD^{+/-}$, whose adaptiveness allowed us to conduct a large empirical analysis. Focusing on a specific index or year can lead to misleading results, as the effectiveness of technical rules appears to be highly volatile.

**Acknowledgements**

We appreciate the comments of Laurent Barras, Campbell R. Harvey, Fearghal Kearney, Neil Kellard, Christopher J. Neely and Guofu Zhou as well as those of participants at the Forecasting Financial Markets (2017) and the Quantitative Finance and Risk Analysis (2017) conferences and seminar participants at the University of Oviedo (2017), Newcastle University (2017), St. Andrews University (2018), University of Glasgow (2018), Cass Business School (2018), University of Glasgow (2019) and University of La Laguna (2019). Any errors which remain are the sole responsibility of the authors.

**Appendices**

**Appendix A. Monte Carlo simulations**

In this appendix, we present supporting evidence of the finite sample performance of the DFDR$^{+/-}$ test using a Monte Carlo experiment. Our main goal is the exploration of the empirical level and power of the test in accurately estimating the proportions of outperforming, underperforming and neutral trading rules. Even though we mainly focus on the FDR rate and its power on the rejection frequency of rules with significant returns (either positive or negative), we also compare it with the power of the FDR procedure presented in Storey (2004).

Before we start running our Monte Carlo simulation, we need to ensure that our experiment correctly embodies the empirical properties of the technical trading strategies employed, such as their time series and cross-sectional dependencies (see also Barras *et al.*, 2010; Hsu *et al.*, 2010; Bajgrowicz and Scaillet, 2012). We have previously demonstrated that our technical trading rules are fully characterized by a weak form of dependence, and this holds especially for those belonging in the same family (e.g. moving averages). This is the main property, and we need to take that into consideration when constructing our experiment. In this way, we can also examine whether our DFDR$^{+/-}$ does indeed have a good response to weak dependence conditions. In order to work in this direction we simultaneously resample matrices of $b \times l$ returns, where $b$ the random block size is consecutive time series observations ($\bar{b} = 10$) under the stationary bootstrap and $l = 21,195$ denotes the trading rules universe as in the empirical exercise. This approach also allows us to preserve the cross-sectional dependencies among the strategies of the same class, while we also preserve autocorrelation every time we apply the same bootstrap replication to all trading rules. For the Monte Carlo experiment, we randomly select the original 155-day sample (i.e. seven months) from July 1, 2013 to February 1, 2014, to simulate our trajectories and we generate the 155-day trajectories for the $l = 21,195$ trading rules as in the empirical exercise. In particular, we employ the stationary bootstrap to create every realized trajectory similar to calculating the *p*-values of the empirical study. We generate 1,000 bootstrap replications of returns, where each replication has similar statistical properties.

In order to obtain the true power of the DFDR$^{+/-}$ test in selecting the proportions of outperforming, underperforming and neutral rules, we need to control these proportions beforehand, likewise observing them a priori. We can then compare them with their corresponding estimations based on the DFDR$^{+/-}$. We adjust 20% of the simulated strategies to outperform the benchmark, 50% to deliver "neutral" returns with no significant performance and 30% to underperform the benchmark during the simulation process. The selected outperforming (underperforming) strategies consist only of a group of neighbouring rules, ranked in terms of highest (lowest) returns in our empirical sample. In this way we avoid having in our groups rules with only slightly different parameters but which at the same time possibly belong to both outperforming and underperforming classes.

In terms of the specific procedure followed, we achieve the control of outperforming, "neutral" and underperforming rules by re-centring the generated returns of each trading rule with its own mean and we utilize that across all five families of rules. This actually leads to all trajectories having almost zero-mean properties while retaining their corresponding, unique standard deviations. We then shift the paths of the outperforming and underperforming rules by some positive and negative value respectively, while keeping each rule's corresponding standard deviation the same. Such a parallel transition does not, however, affect the empirical properties of the paths, other than the mean (see Paparoditis and Politis, 2003). The notion is to construct the trajectories of different strategies in such a way as to exactly acquire the same, positive Sharpe ratio for all outperforming rules and the same negative Sharpe ratio for all underperforming rules. We specify both chosen positive and negative Sharpe ratios in advance[23].

As for the target Sharpe ratios employed for shifting the paths of outperforming and underperforming strategies, we follow the study of Bajgrowicz and Scaillet (2012) and select Sharpe ratios closely related to those obtained in our empirical exercise. In particular, we set three specific targets of positive Sharpe ratios for our outperforming rules, i.e., 2, 3, 4; and three specific targets of negative Sharpe ratios for our underperforming rules, i.e., -2, -3, -4. All of them correspond to annualized Sharpe ratios, just like those calculated from the daily returns of each strategy. We then consider pairs of the Sharpe ratios above in order to adjust the outperforming and underperforming rules, while shifting their trajectories towards the target. Take the (2, -2) pair for example. We design 20% of the rules to yield an equal Sharpe ratio of 2 (i.e. outperforming) and likewise all 30% of the rules share an equal Sharpe ratio of -2 (i.e. underperforming). The remaining 50% of our rules' universe show zero performance. This results in nine possible combinations of positive and negative Sharpe ratio pairs representing fixed alternative hypotheses against the null of a Sharpe ratio being equal to zero. The above levels seem to match our historical sample results since we obtain positive annualized Sharpe ratios up to 4 for the best-performing strategies and negative annualized Sharpe ratios down to

---

[23] We multiply the corresponding standard deviation of each rule by the pre-specified Sharpe ratio and we add the calculated value to each data point so that the mean for the rule becomes SR sigma.

-4 for the worst-performing ones. However, the outperformance versus underperformance pair of (2, -2) still portrays quite a challenging set-up for our portfolio construction method.

We present the results of our Monte Carlo experiments in Tables A.1-A.3 below. Table A.1 displays the annualized mean excess return quartiles for the controlled outperforming and underperforming technical trading rules based on the 1,000 Monte Carlo replications for the nine combinations of Sharpe ratios examined.

[Table A.1 here]

In general, the annualized mean returns we obtain seem quite analogous to their corresponding Sharpe ratio levels, either positive or negative.

Focusing on the estimation power of the DFDR$^{+/-}$ approach, Table A.2 presents the estimates for the proportions of outperforming ($\widehat{\pi_A^+}$), underperformming ($\widehat{\pi_A^-}$) and neutral ($\widehat{\pi_0}$) strategies under the Sharpe ratio metric and for the nine possible Sharpe ratio pairs. It also reports the success of the estimators in tracking the actual proportions of outperforming ($\pi_A^+ = 20\%$), underperforming ($\pi_A^- = 30\%$), and neutral ($\pi_0 = 50\%$) trading rules. Once again, we apply the "point estimates method" of Storey et al. (2004) to the DFDR $^{+/-}$ test to obtain the estimators of these proportions based on the Monte Carlo results. This time we keep the cut-off threshold fixed to $\gamma^* = 0.4$, as at this point $\widehat{\pi_A^+}$ and $\widehat{\pi_A^-}$ become constant (see Barras et al. 2010). In other words as $\gamma$ increases up to an adequate enough value, $\widehat{\pi_A^+}$ and $\widehat{\pi_A^-}$ include both genuine and false selections of trading rules representing the total number of outperforming and underperforming rules respectively.

[Table A.2 here]

Our DFDR$^{+/-}$ approach seems to provide quite robust estimators for the outperforming, underperforming and neutral proportions of technical trading rules, with only small deviations from their true corresponding levels. For instance, looking at the (3, -3) Sharpe ratios pair, the estimator for the outperforming rules (i.e. $\widehat{\pi_A^+}$), is 15.23%, the relevant estimator for underperforming rules (i.e. $\widehat{\pi_A^-}$) is 27.68% and that for neutral rules (i.e. $\widehat{\pi_0}$) is 57.09%, which are quite close to their true levels of 20%, 30% and 50% respectively. This clearly highlights the power of our method in accurately identifying the true proportions of outperforming, underperforming and neutral rules in the entire population.

Finally, we present in Table A.3 the performance of constructed portfolios of outperforming rules under the DFDR$^+$ approach based on the Monte Carlo simulation and for each of the nine Sharpe ratio combinations. We control the DFDR$^+$ at a prespecified level similar to our empirical exercise. For instance, we build two different types of DFDR$^+$ portfolios by setting the targets of erroneous selections at 10% and 20% respectively. In terms of performance and power, the table reports the actual false discovery rate achieved (FDR$^+$) in comparison with its fixed level adjusted in advance (i.e. 10% and 20%), the proportions of genuinely best-performing rules over the total number of outperforming rules denoted as "power", and the absolute number of genuinely best-performing trading rules as "portfolio size"[24]. To reflect the marginal contribution of the proposed method, we compare our results with the FDR procedure of Storey *et al.* (2004), while controlling the FDR at the target levels of 10% and 20% respectively.

[Table A.3 here]

The findings of Table A.3 reveal that the DFDR$^{+/-}$ approach is superior in terms of finite sample power to the more conservative FDR approaches such as the one in Storey *et al.* (2004). Specifically, the DFDR$^{+/-}$ reports robust power in rules selection and portfolio size, while it closely tracks the actual false discovery rate across all conditions and Sharpe ratio pairs. For example, consider again the (3, -3) Sharpe ratios pair, where the 10%-DFDR$^+$ portfolio efficiently converges to its FDR rate at 8% and successfully discovers on average 64.74% of the best-performing rules. On the other hand, the corresponding 10%-FDR portfolio discovers only 30.27% of the best-performing rules on average, while it meets its target rate only at 7.3%. When it comes to the size of portfolios, the 10%-DFDR$^+$ outstandingly outperforms the 10%-FDR approach by sufficiently selecting 3,048 rules, while the benchmark method detects only 1907. Increasing the target rate of the FDR to 20% does not affect the patterns since it improves the power of selection to 39.86% while the portfolio size is not affected (1,907). The 20%-DFDR$^+$ though, performs even better by detecting on average 66.3% of the outperforming rules, and forms a portfolio of 3,322 trading rules. In terms of target rate, the 20%-DFDR$^+$ portfolio falls below 20% and achieves an FDR$^+$ of 10.59%. Asymptotic theory is the most possible reason for this outcome, but the 20%-DFDR$^+$ portfolio is still able to successfully deal with data snooping bias as seen above. Overall, our Monte Carlo experiments undoubtedly reveal that the DFDR$^{+/-}$ method has greater power when compared with conservative FDR methods, such as the Storey *et al.* (2004) procedure.

---

[24] We compute the actual false discovery rate (FDR$^+$) by replacing the actual proportion of neutral trading rules (i.e. instead of the estimated one (i.e., ) in .

**Appendix B. IS Performance, Disaggregation Analysis and Robustness Exercise**

We repeat all the exercises by setting the look-back period to one year and the OOS to 3, 6 and 9 months respectively this time. Figure B.1 and tables B.1 to B.5 report all relevant results for IS performance, break-even transaction costs and OOS performance.

Tables B.6 to B.10 present the exact contribution, in percentage terms, of every class of rules to the overall universe of survivors every year, as those have been presented on average in Table 4 in section 5.1.

**B.1. IS Performance**

As about IS performance, Table B.2 presents the annualized returns and Sharpe ratios of significant rules after one-way transaction costs during a look-back period of one year, similar to Table 3 in the IS performance section.

[Table B.2 here]

Comparing the corresponding Tables 3 and B.2, we conclude that when using one year as our look-back period, the performance of trading rules is somewhat better in terms both of excess mean return and Sharpe ratio criteria. Of course, this was somehow anticipated since the bigger the sample period, the more the technical trading rules are exposed to fluctuations and market risks in general, leading to lower performance most of the time. Realizing higher IS returns in smaller sample horizons is a common phenomenon not only for the relevant literature but also for trading desks.

**B.2 Disaggregation Analysis**

We perform a disaggregation analysis for the five families of technical trading rules separately and we present the percentage profitability of every single family among the DFDR$^{+/-}$ procedure survivors in Tables B.3-B.7.

[Table B.3 here]

[Table B.4 here]

[Table B.5 here]

[Table B.6 here]

[Table B.7 here]

The RSI and support and resistance rules seem to perform better in recent periods (i.e., 2012-2015), reporting a contribution in the portfolios of significant rules, while only few of them survive over the earlier periods. Additionally, the filter rules' performance varies over the years, with 2009 and 2010 being the best years in terms of profitability. The same finding seems to hold for channel breakout rules, however they seem to contribute more to the overall portfolio performance up to 2010. Finally, moving averages retain the biggest proportion in the portfolios' profitability over the years, with the first half of the examined period (i.e., 2006-2011) being their best in terms of performance.

**B.3 Break-even Transaction Costs**

In terms of break-even transaction costs, Figure B.1 demonstrates the corresponding size of average break-even transaction costs per year for the best-performing technical trading rule selected on a monthly OOS basis and for each index separately. However, this time the IS period covers one year.

[Figure B.1 here]

Once again frontier and emerging markets (i.e. Brazil, Morocco, Russia) report the highest break-even transaction costs over the first four years in general. In addition, the UK surprisingly achieves one of the highest break-even costs in 2009. When it comes to the rest of the years, until the very recent periods there is a considerable decay of these costs, similar to Figure 1, in which a two-year IS is employed. However, this is not always the case for all markets examined. For example, for some emerging and frontier countries, such as Russia and Estonia, the break-even transaction costs seem to recover in 2015 and 2013 respectively. Overall, the break-even transaction costs as well as their evident cyclicality are not that high here compared to those under the two-year IS as presented in Figure 1.

**B.4 OOS Performance**

However, what matters most for a technical trader is the OOS simulation findings rather than the IS ones. Tables B.8-B.10 correspond to a one-year look-back period as IS while considering the OOS periods of one, three and six months respectively.

[Table B.8 here]

[Table B.9 here]

[Table B.10 here]

In this case, we observe opposing evidence with regards to the IS period chosen each time. For example, comparing Tables 4 – 6 corresponding to the IS period of two years and the same OOS periods with Tables B.8 – B.10, the results provided by the first approach are more favourable. We may attribute these findings to the fact that including more information (a larger historical sample) when searching for a predictive technical trading rule IS, results in a better performance OOS. Moreover, we can also explain the above findings in terms of technical trading rules' specific characteristics and parameterizations. For instance, we utilize technical trading rules, even of the same family, whose lagged values span from one day up to one year (e.g. a double moving average of two and five days respectively; a double moving average of 150 and 250 days respectively). This means that they both need different time periods in order to capture all the available market trends, momentum or reversals. Choosing a small IS period (i.e. one year) might provide enough information for trading rules utilizing short periods of previous market returns (i.e. a double moving average of two and five days) but not enough inputs for trading rules looking back at longer periods of market movements (i.e. a double moving average of 150 and 250 days respectively). Hence, in our opinion considering a sufficient enough horizon based on a strategy's properties, while setting an optimal IS/OOS ratio, is equally important for the selection of the best predictive rule.

We investigate further the above optimality in the IS and OOS ratio with respect to sample periods chosen by looking at the corresponding performances of the significant technical trading rules over the three different OOS periods (i.e. one, three and six months) examined and the IS period of one year in this appendix. In terms of average annual performance of all markets considered (i.e. last row), Tables B.8 – B.10 reveal specific patterns in OOS excess profitability of technical trading rules according to both mean return and Sharpe ratio metrics. Specifically, from 2006 to 2009 employing the short OOS period of one month achieves higher mean returns as well as Sharpe ratios compared to the longer periods used (i.e. three and six months), which display a decay as the OOS periods becomes larger during these years. On the contrary, there is a turning point in this phenomenon for the period 2010-2012. The longer the OOS period, the greater the mean return and Sharpe ratio. Despite that, we must note that both metrics appear negative during these years. For the rest of the years (i.e. 2013-2015), technical trading rules seem to perform better using the OOS period of one month, even yielding positive metrics in 2015. In general, profitability diminishes as we approach the most recent periods for all OOS periods and across all markets considered. This evidence is consistent with the general findings presented in Section 6.1 when a two-year IS horizon was considered.

When it comes to each market's average performance over the full ten-year period (i.e. last column) the picture is quite different. There is no clear evidence in support of a specific OOS period in general and sometimes both performance metrics employed provide contradictory results. We conclude that the most suitable OOS horizon depends on the specific market exploited. For advanced markets, the performance of trading rules seems to improve according to the Sharpe ratio as we expand the OOS period, but this is not the case when the mean excess return is adopted as the performance criterion. As for the emerging and frontier markets, results provide the opposite result, in which the shorter the OOS periods employed, the better the technical trading rules performance, which is consistent with both mean return and Sharpe ratio. Despite that, we must also mention that technical trading rules underperform the benchmark most of the time, especially in the advanced markets, which once again justifies the use of an IS period of two years.

**Figures**

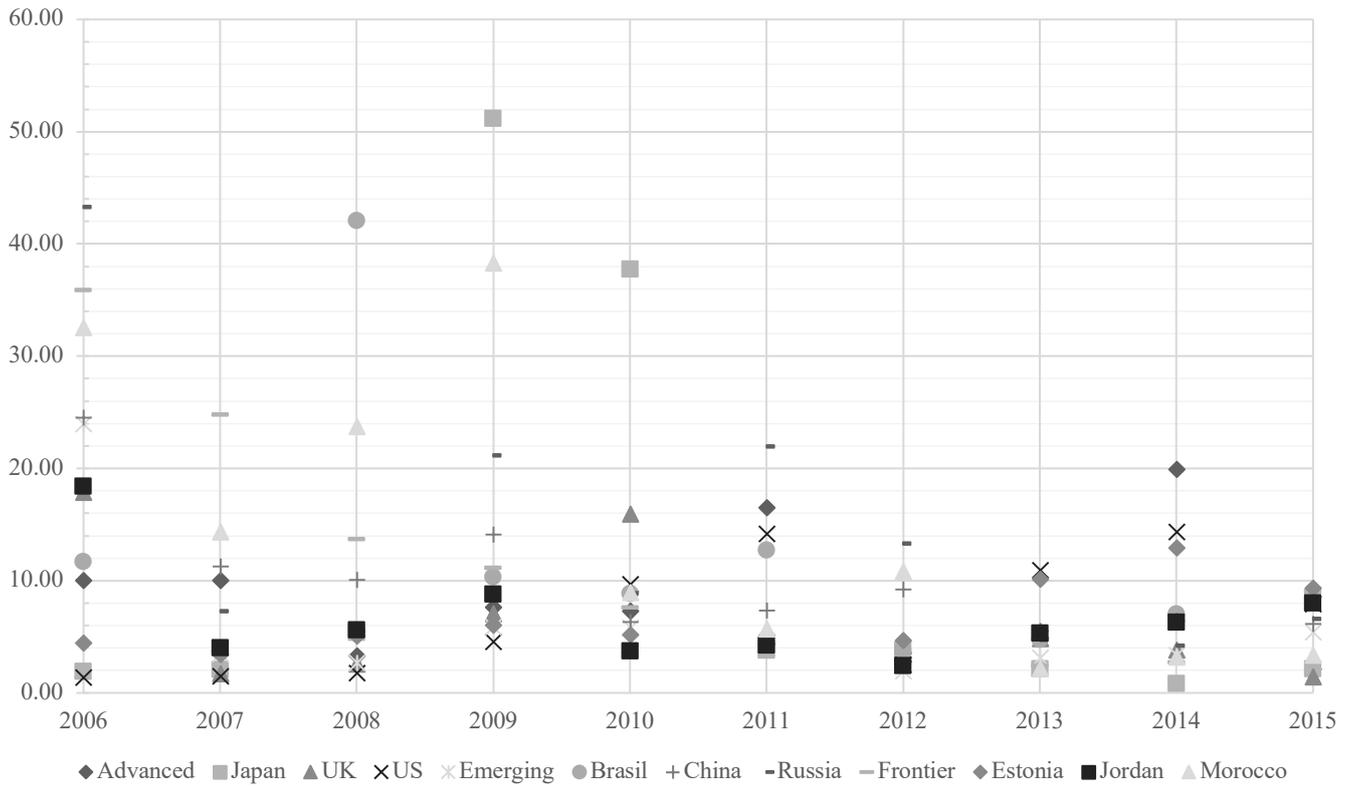

◆Advanced  ■Japan  ▲UK  ✕US  ✳Emerging  ●Brasil  +China  −Russia  —Frontier  ◆Estonia  ■Jordan  ▲Morocco

**Figure 1. Break-even cost for the top performing survivor of the DFDR [+] procedure (IS 2 Years)**
The values are in percentages and calculated as the transaction cost that sets the excess return to zero over the period under study. The IS period is set at two years, while the same results for IS of one year is available in Figure A.1. The values are calculated by repeating the procedure at the start of each month and averaging over 12 months.

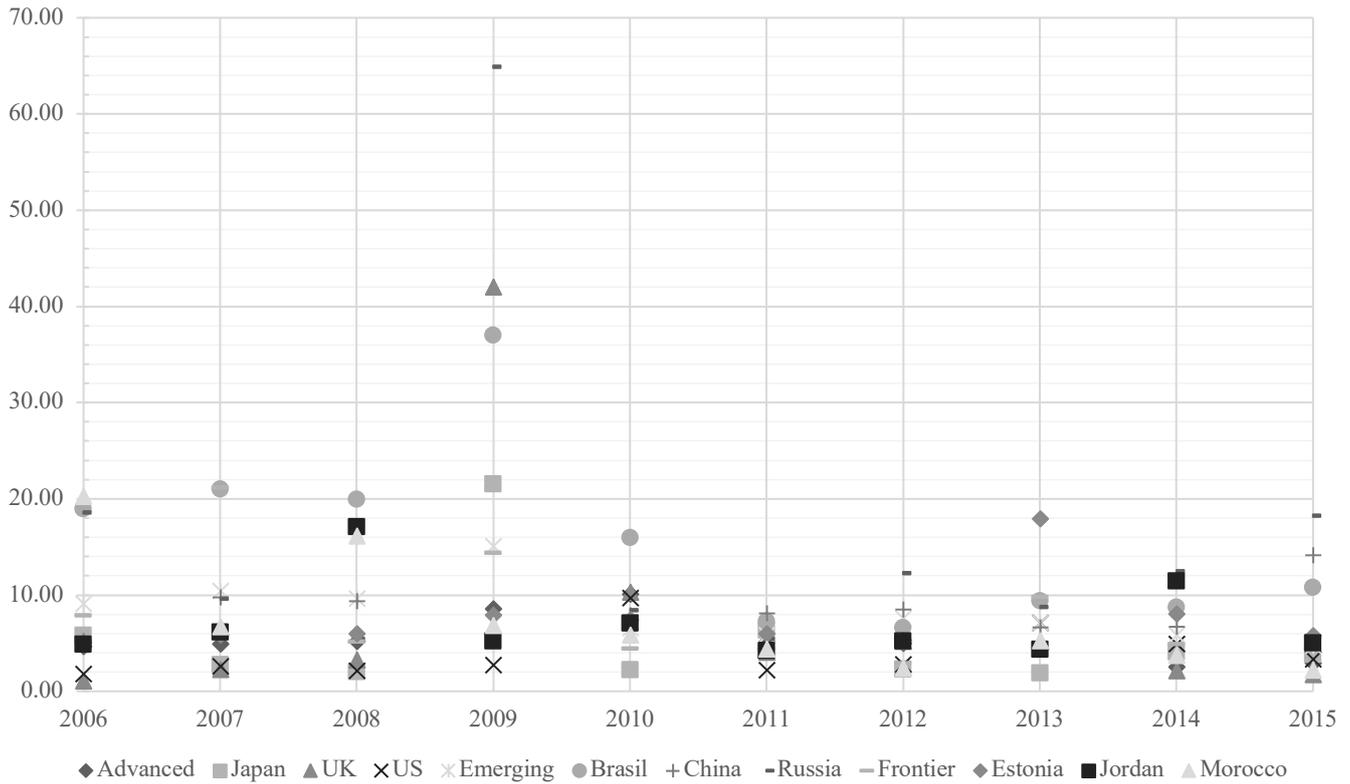

**Figure B.1. Break-even cost for the top performing survivor of the DFDR+ procedure (IS 1 Year)**
The values are in percentages and calculated as the transaction cost that sets the excess return to zero over the period under study. The IS period is set at one year. The values are calculated by repeating the procedure at the start of each month and averaging over 12 months.

# Tables

**Table 1**
**Summary statistics of the daily return series under study (12 MSCI indices and the federal funds rate).**

| Market | Mean (bp) | Max (%) | Min (%) | Std. dev. (%) | Kurtosis | Skewness | First AC (significance) |
|---|---|---|---|---|---|---|---|
| **Advanced** | 1.55 | 9.10 | -7.33 | 1.02 | 12.86 | -0.50 | 0.12 (*) |
| US | 1.45 | 11.04 | -9.51 | 1.18 | 15.11 | -0.36 | -0.10 (*) |
| UK | 1.60 | 17.32 | -36.26 | 1.29 | 212.22 | -6.62 | 0.01 |
| Japan | 2.66 | 12.77 | -20.75 | 1.27 | 62.62 | -2.12 | -0.07 |
| **Emerging** | 1.91 | 10.07 | -9.99 | 1.27 | 11.38 | -0.49 | 0.22 (*) |
| Russia | 1.83 | 42.37 | -58.10 | 2.35 | 172.93 | -2.26 | 0.02 |
| China | 3.32 | 14.05 | -12.84 | 1.74 | 10.10 | -0.04 | 0.03 (*) |
| Brazil | 3.59 | 37.69 | -46.23 | 2.19 | 109.54 | -0.39 | 0.02 |
| **Frontier** | 1.95 | 12.54 | -9.32 | 1.62 | 8.74 | 0.20 | 0.06 (*) |
| Estonia | 2.96 | 5.50 | -7.70 | 1.06 | 6.47 | -0.13 | 0.16 (*) |
| Morocco | 0.99 | 5.69 | -9.07 | 0.83 | 15.56 | -1.38 | 0.26 (*) |
| Jordan | 1.21 | 7.82 | -9.08 | 1.10 | 13.03 | -0.71 | 0.07 (*) |
| **Federal funds rate** | 0.53 | 0.02 | 0.00 | 0.01 | 2.78 | 1.18 | 1.00 (*) |

The mean daily returns are reported in basis points (bp). Maximum, minimum and standard deviation are presented in percentages (%). The last column reports the first-order autocorrelation coefficients. Coefficients notated with (*) are significant at 1% (*) level for the Ljung-Box Q statistic. The study period for all time series is 01/01/2004 to 31/12/2016.

**Table 2**
**Percentage and standard deviation of the DFDR$^{+/-}$ procedure survivors (IS 2 years).**

| Market | 2006 | 2007 | 2008 | 2009 | 2010 | 2011 | 2012 | 2013 | 2014 | 2015 | Average |
|---|---|---|---|---|---|---|---|---|---|---|---|
| **Advanced** | 0.34 (0.97) | 1.16 (1.87) | 0.43 (1.03) | 2.39 (1.78) | 2.84 (1.56) | 3.52 (7.03) | 0.24 (0.06) | 3.50 (7.02) | 16.24 (11.37) | 0.33 (0.15) | **3.10 (3.28)** |
| US | 0.01 (0.00) | 0.02 (0.01) | 0.16 (0.32) | 0.91 (1.12) | 1.54 (0.94) | 8.33 (9.49) | 0.28 (0.07) | 7.75 (10.55) | 31.49 (3.55) | 19.58 (13.81) | **7.01 (3.99)** |
| UK | 20.08 (9.08) | 15.52 (10.26) | 0.32 (0.38) | 12.97 (10.21) | 20.89 (6.96) | 9.41 (9.88) | 0.18 (0.04) | 0.15 (0.06) | 8.72 (11.71) | 0.29 (0.09) | **8.85 (5.87)** |
| Japan | 3.59 (6.56) | 0.11 (0.04) | 0.69 (1.49) | 3.20 (1.32) | 2.58 (0.29) | 1.09 (1.40) | 0.29 (0.05) | 0.27 (0.07) | 0.18 (0.03) | 0.16 (0.04) | **1.22 (1.13)** |
| **Emerging** | 2.87 (5.39) | 1.26 (1.19) | 1.12 (1.46) | 3.35 (0.97) | 6.04 (4.46) | 9.01 (10.21) | 0.36 (0.09) | 0.39 (0.10) | 0.19 (0.08) | 0.24 (0.12) | **2.48 (2.41)** |
| Russia | 14.39 (7.39) | 16.22 (9.55) | 1.77 (3.48) | 25.64 (7.84) | 17.61 (4.78) | 10.90 (10.65) | 0.28 (0.10) | 0.44 (0.11) | 0.58 (0.14) | 0.90 (0.30) | **8.87 (4.43)** |
| China | 2.59 (3.38) | 32.32 (15.03) | 5.14 (9.68) | 0.93 (0.56) | 3.72 (6.77) | 3.59 (7.16) | 0.28 (0.01) | 0.27 (0.07) | 0.20 (0.09) | 0.81 (0.52) | **4.99 (4.33)** |
| Brazil | 22.52 (13.6) | 8.62 (7.28) | 8.84 (8.44) | 17.79 (5.10) | 20.84 (5.19) | 8.71 (9.74) | 0.09 (0.07) | 0.27 (0.35) | 1.02 (0.38) | 0.32 (0.14) | **8.9 (5.03)** |
| **Frontier** | 14.14 (12.40) | 1.10 (0.45) | 3.85 (7.50) | 29.22 (7.27) | 25.26 (9.00) | 7.24 (9.64) | 0.37 (0.19) | 0.47 (0.23) | 4.39 (7.36) | 1.49 (1.26) | **8.75 (5.53)** |
| Estonia | 17.37 (16.94) | 0.35 (0.71) | 4.34 (7.12) | 7.91 (2.73) | 7.52 (4.13) | 10.23 (10.14) | 0.19 (0.05) | 4.35 (7.18) | 8.86 (6.51) | 0.66 (1.26) | **6.18 (5.68)** |
| Morocco | 7.36 (8.27) | 26.96 (8.76) | 17.24 (9.58) | 4.82 (2.90) | 0.64 (0.63) | 0.15 (0.06) | 0.34 (0.34) | 0.65 (0.62) | 0.15 (0.05) | 0.22 (0.10) | **5.85 (3.13)** |
| Jordan | 20.26 (11.4) | 1.57 (2.22) | 1.77 (1.67) | 4.27 (1.34) | 1.52 (0.62) | 0.67 (0.84) | 0.21 (0.05) | 0.09 (0.03) | 0.11 (0.03) | 0.18 (0.07) | **3.06 (1.83)** |
| **Average** | **10.46 (7.95)** | **8.77 (4.78)** | **3.81 (4.35)** | **9.45 (3.60)** | **9.25 (3.78)** | **6.07 (7.19)** | **0.26 (0.09)** | **1.55 (2.20)** | **6.01 (3.44)** | **2.10 (1.49)** | **5.77 (3.89)** |

This table reports the percentage and standard deviations of the survivor rules adjusted by the total number of rules. For example, in 2006 for the advanced market, the average number of surviving rules is 72 (0.0034*21195) and their standard deviation is 206 (0.0097*21195). The average is estimated from the twelve portfolios whose OOS is in 2006. The first portfolio's IS runs from 01/01/2004-31/12/2005 and the remaining eleven are calculated by rolling forward the IS by one month.

**Table 3**
**Annualized Returns and Sharpe Ratios after Transaction Costs (IS 2 Years)**

| Market | 2006 | 2007 | 2008 | 2009 | 2010 | 2011 | 2012 | 2013 | 2014 | 2015 | Average |
|---|---|---|---|---|---|---|---|---|---|---|---|
| **Advanced** | 10.67% (1.62) | 11.87% (1.44) | 14.93% (1.39) | 27.00% (1.66) | 24.73% (1.62) | 8.65% (1.79) | 8.06% (1.75) | 7.92% (1.65) | 9.36% (1.57) | 5.64% (2.39) | **12.88% (1.69)** |
| US | 8.65% (0.95) | 10.41% (1.11) | 14.36% (1.11) | 25.44% (1.19) | 24.93% (1.20) | 9.54% (1.38) | 8.45% (1.50) | 10.73% (1.51) | 10.55% (1.48) | 7.50% (1.24) | **13.05% (1.27)** |
| UK | 10.13% (1.21) | 11.00% (1.21) | 15.50% (0.90) | 23.53% (0.99) | 28.08% (1.16) | 11.32% (1.25) | 4.89% (2.19) | 5.48% (1.33) | 11.53% (1.52) | 8.02% (1.99) | **12.95% (1.38)** |
| Japan | 14.23% (1.02) | 13.37% (0.98) | 9.13% (0.73) | 19.31% (1.10) | 18.57% (1.18) | 7.24% (1.17) | 7.05% (1.38) | 5.73% (1.64) | 3.70% (1.64) | 3.14% (1.83) | **10.15% (1.27)** |
| **Emerging** | 19.92% (2.35) | 22.49% (2.09) | 26.91% (1.84) | 37.28% (2.04) | 34.96% (1.98) | 12.36% (1.88) | 12.61% (2.08) | 12.86% (2.09) | 8.83% (2.11) | 6.25% (1.89) | **19.45% (2.03)** |
| Russia | 22.47% (0.95) | 22.45% (0.98) | 16.30% (0.70) | 46.59% (1.18) | 47.53% (1.30) | 15.61% (1.15) | 14.90% (1.74) | 17.32% (1.75) | 11.51% (1.76) | 22.86% (1.65) | **23.76% (1.32)** |
| China | 23.05% (1.65) | 24.14% (1.73) | 35.52% (1.68) | 43.71% (1.52) | 32.85% (1.37) | 8.73% (1.37) | 5.51% (2.33) | 8.40% (1.90) | 12.01% (1.75) | 14.51% (1.65) | **20.84% (1.70)** |
| Brazil | 24.90% (1.35) | 27.61% (1.06) | 30.54% (1.02) | 37.87% (1.25) | 35.75% (1.22) | 11.15% (1.17) | 11.41% (1.98) | 15.05% (1.83) | 13.20% (1.85) | 17.85% (1.48) | **22.53% (1.42)** |
| **Frontier** | 16.79% (2.60) | 17.46% (2.33) | 20.21% (1.95) | 29.42% (2.06) | 28.96% (2.24) | 12.98% (2.39) | 10.32% (2.17) | 10.23% (2.24) | 9.39% (2.27) | 10.49% (2.02) | **16.62% (2.23)** |
| Estonia | 18.07% (1.81) | 20.66% (1.64) | 29.00% (1.77) | 42.44% (1.73) | 40.75% (1.78) | 18.93% (1.47) | 13.62% (1.77) | 12.09% (1.58) | 13.37% (1.38) | 10.94% (1.37) | **21.99% (1.63)** |
| Morocco | 21.44% (2.07) | 21.55% (1.91) | 24.35% (1.71) | 27.58% (1.60) | 16.92% (1.54) | 4.60% (1.99) | 10.91% (1.17) | 12.83% (1.21) | 4.86% (1.95) | 3.47% (1.34) | **14.85% (1.65)** |
| Jordan | 33.77% (1.97) | 25.68% (1.56) | 21.99% (1.59) | 28.3% (1.59) | 22.09% (1.43) | 12.05% (1.79) | 9.5% (1.83) | 4.60% (1.91) | 6.92% (1.52) | 5.79% (1.55) | **17.07% (1.67)** |
| **Average** | **18.67% (1.63)** | **19.06% (1.50)** | **21.56% (1.37)** | **32.37% (1.49)** | **29.68% (1.50)** | **11.10% (1.57)** | **9.77% (1.82)** | **10.27% (1.72)** | **9.60% (1.73)** | **9.70% (1.70)** | **17.18% (1.60)** |

This table reports the average IS annualized returns and Sharpe ratios of twelve portfolios for two years of IS after transaction costs (rolling forward by one month). For example, the 10.67% annualized return of the advanced markets (2006) is calculated as the average IS annualized return of twelve portfolios. The first portfolio's IS return is calculated over the period of 01/01/2004-31/12/2005. The remaining eleven are calculated by rolling forward the IS by one month. The same logic applies for the Sharpe ratios. The last column and row present the average performance per market across all years and per year respectively.

**Table 4**
**Percentage decomposition of the DFDR$^{+/-}$ procedure survivors in classes of technical trading rules (IS 2 Years)**

| Market | Average of RSI% | Average of FR% | Average of MA% | Average of SR% | Average of CB% |
|---|---|---|---|---|---|
| **Advanced** | 4.97% | 8.54% | 65.78% | 7.30% | 13.40% |
| US | 0.85% | 10.84% | 61.46% | 10.07% | 16.78% |
| UK | 10.09% | 11.32% | 60.81% | 5.79% | 12.00% |
| Japan | 8.31% | 8.93% | 39.10% | 22.42% | 21.25% |
| **Emerging** | 6.07% | 21.97% | 42.98% | 13.98% | 15.00% |
| Russia | 3.57% | 18.28% | 48.54% | 19.32% | 10.29% |
| China | 3.69% | 8.71% | 63.61% | 16.77% | 7.22% |
| Brazil | 7.09% | 10.96% | 56.41% | 13.31% | 12.23% |
| **Frontier** | 3.19% | 16.79% | 52.88% | 14.23% | 12.90% |
| Estonia | 1.27% | 7.65% | 69.34% | 12.77% | 8.96% |
| Morocco | 9.81% | 6.12% | 59.34% | 18.38% | 6.35% |
| Jordan | 16.93% | 9.26% | 54.13% | 14.15% | 5.54% |
| **Grand Total** | **6.32%** | **11.61%** | **56.20%** | **14.04%** | **11.83%** |

This table reports the decomposition (in percentage terms) of the DFDR$^{+/-}$ procedure survivors in every single family of technical trading rules as an average across all years examined. For example, for the advanced market, 4.97% of the survivors belongs to RSI rules, 8.54% to filter rules, 65.78% to moving averages, 7.3% to support and resistance and 13.4% to channel breakouts on average for the period 2006-2015. For every year, the IS portfolio runs for two years and the remaining eleven are calculated by rolling forward the IS by one month.

**Table 5**
**Annualized Returns and Sharpe Ratios after Transaction Costs (IS 2 Years and OOS 1 Month)**

| Market | 2006 | 2007 | 2008 | 2009 | 2010 | 2011 | 2012 | 2013 | 2014 | 2015 | Average |
|---|---|---|---|---|---|---|---|---|---|---|---|
| **Advanced** | -3.66% | -4.72% | 10.63% | 4.14% | -17.79% | -1.92% | -3.23% | 0.35% | -2.07% | -1.43% | -1.97% |
|  | (-0.54) | (-0.43) | (0.41) | (0.25) | (-2.91) | (-0.5) | (-1.27) | (0.13) | (-0.42) | (-0.64) | (-0.59) |
| US | 1.81% | -6.16% | 22.72% | -3.64% | -14.07% | -1.65% | -1.37% | 7.02% | 5.73% | -5.28% | 0.51% |
|  | (0.2) | (-0.45) | (0.77) | (-0.19) | (-1.49) | (-0.31) | (-0.5) | (1.31) | (0.77) | (-0.71) | (-0.06) |
| UK | 15.94% | -3.14% | 16.5% | 19.58% | -12.73% | -4.12% | -3.83% | 1.67% | -7.09% | -2.8% | 2% |
|  | (1.55) | (-0.27) | (0.41) | (1.16) | (-1.47) | (-0.58) | (-3.7) | (0.67) | (-0.97) | (-0.45) | (-0.36) |
| Japan | -11.33% | -8.68% | 41.53% | 11.47% | -2.11% | -7.87% | -3.3% | -3.95% | -5.75% | -2.98% | 0.7% |
|  | (-0.76) | (-1.8) | (1.4) | (0.51) | (-0.37) | (-1.62) | (-1.31) | (-1.03) | (-2.27) | (-1.13) | (-0.84) |
| **Emerging** | 1.75% | -1.93% | 43.52% | 4.25% | -7.22% | -3.14% | -0.93% | -1.71% | -3.71% | -1.16% | 2.97% |
|  | (0.15) | (-0.15) | (1.1) | (0.29) | (-1.09) | (-0.52) | (-0.23) | (-0.55) | (-1.28) | (-0.31) | (-0.26) |
| Russia | 45.77% | -9% | 45.64% | 13.09% | -20.64% | -9.9% | -2.42% | -4.63% | 6.92% | -12.48% | 5.24% |
|  | (1.25) | (-1.33) | (1.01) | (0.47) | (-2.1) | (-1.02) | (-0.44) | (-0.92) | (0.44) | (-0.98) | (-0.36) |
| China | 59.59% | 29.16% | -2.1% | -19.48% | -8.93% | -10.76% | -0.29% | -3.54% | -4.51% | 11.46% | 5.06% |
|  | (2.62) | (1.17) | (-0.05) | (-1.04) | (-1.09) | (-1.4) | (-0.64) | (-1.42) | (-0.66) | (0.82) | (-0.17) |
| Brazil | 7.53% | 67.85% | 48.86% | 15.4% | -15.81% | -5.12% | -7.98% | -3.03% | -7.35% | -1.07% | 9.93% |
|  | (0.28) | (1.2) | (0.9) | (0.68) | (-1.67) | (-0.73) | (-2.42) | (-0.49) | (-0.6) | (-0.07) | (-0.29) |
| **Frontier** | -11.64% | 14.24% | 64.67% | 11.12% | 0.26% | -12.4% | -0.27% | 4.2% | 7.75% | 2.33% | 8.03% |
|  | (-2.04) | (1.44) | (2.24) | (1.04) | (0.06) | (-3.55) | (-0.09) | (0.98) | (1.39) | (0.32) | (0.18) |
| Estonia | -4.93% | -7.76% | 65.08% | 2.83% | 6.06% | -24.12% | 6.26% | -4.29% | 12.92% | -13.23% | 3.88% |
|  | (-0.62) | (-0.45) | (1.46) | (0.1) | (0.3) | (-2.6) | (0.98) | (-0.69) | (1.23) | (-2.11) | (-0.24) |
| Morocco | 34% | 21.97% | 25.32% | 1.5% | -10.22% | -2.49% | 1.19% | -4.9% | -0.64% | -0.36% | 6.54% |
|  | (1.87) | (1.65) | (1.2) | (0.1) | (-2.17) | (-0.87) | (0.1) | (-0.48) | (-0.61) | (-0.17) | (0.06) |
| Jordan | -0.62% | -3.47% | 37.89% | -9.46% | -0.12% | -0.79% | -4.59% | -2.27% | -4.63% | -1.32% | 1.06% |
|  | (-0.04) | (-0.43) | (1.32) | (-0.81) | (-0.02) | (-0.13) | (-1.55) | (-1.03) | (-1.23) | (-0.37) | (-0.43) |
| **Average** | 11.18% | 7.36% | 35.02% | 4.23% | -8.61% | -7.02% | -1.73% | -1.26% | -0.2% | -2.36% | 3.66% |
|  | (0.33) | (0.01) | (1.02) | (0.21) | (-1.17) | (-1.15) | (-0.92) | (-0.29) | (-0.35) | (-0.48) | (-0.28) |

This table reports the average OOS annualized returns and Sharpe ratios of twelve portfolios for two years of IS and one month of OOS after transaction costs (rolling forward by one month). For example, the -3.66% annualized return of the advanced markets (2006) is calculated as the average OOS annualized return of twelve portfolios. The first portfolio's OOS return is calculated over January 2006 using as IS the period 01/01//2004-31/12/2005. The remaining eleven OOS returns are calculated by rolling forward the IS by one month. The same logic applies for the Sharpe ratios. The last column and row present the average performance per market across all years and per year respectively.

**Table 6**
**Annualized Returns and Sharpe Ratios after Transaction Costs (IS 2 Years and OOS 3 Months)**

| Market | 2006 | 2007 | 2008 | 2009 | 2010 | 2011 | 2012 | 2013 | 2014 | 2015 | Average |
|---|---|---|---|---|---|---|---|---|---|---|---|
| **Advanced** | -4.54% (-0.62) | -3.26% (-0.27) | 24.85% (0.99) | -5.42% (-0.4) | -13.37% (-2.23) | -2.42% (-0.53) | -2.91% (-1.22) | -1.35% (-0.46) | 0.20% (0.04) | -2.24% (-1.06) | **-1.05% (-0.57)** |
| US | 4.28% (0.48) | -4.93% (-0.36) | 34.43% (1.12) | -14.00% (-0.89) | -9.63% (-1.10) | -5.86% (-0.86) | -0.98% (-0.41) | 5.41% (1.03) | 6.43% (0.86) | -4.84% (-0.68) | **1.03% (-0.08)** |
| UK | 14.9% (1.48) | -13.25% (-0.88) | 35.36% (0.79) | 9.70% (0.67) | -9.29% (-1.03) | -5.68% (-0.8) | -2.07% (-2.10) | 0.82% (0.31) | -5.53% (-0.81) | -2.75% (-0.46) | **2.22% (-0.28)** |
| Japan | -16.79% (-0.96) | -9.00% (-1.81) | 29.70% (1.03) | 4.98% (0.23) | -1.76% (-0.35) | -5.78% (-1.53) | -3.03% (-1.25) | -2.77% (-0.86) | -2.84% (-1.70) | -3.67% (-1.49) | **-1.10% (-0.87)** |
| **Emerging** | -4.31% (-0.42) | -2.01% (-0.15) | 15.33% (0.49) | -1.19% (-0.10) | -5.19% (-0.79) | -0.19% (-0.03) | -1.92% (-0.52) | -0.38% (-0.11) | -4.35% (-1.74) | -1.54% (-0.4) | **-0.57% (-0.38)** |
| Russia | 37.24% (1.11) | -6.90% (-0.54) | 31.94% (0.95) | 3.01% (0.13) | -17.55% (-1.79) | -7.51% (-0.82) | -0.09% (-0.02) | -5.58% (-1.08) | 0.52% (0.04) | -12.47% (-1.08) | **2.26% (-0.31)** |
| China | 35.48% (1.70) | 29.13% (1.20) | 3.30% (0.09) | -8.37% (-0.51) | -6.97% (-0.89) | -4.61% (-0.71) | -0.42% (-0.87) | -3.18% (-1.24) | -6.54% (-1.04) | 3.89% (0.32) | **4.17% (-0.19)** |
| Brazil | 1.52% (0.06) | 50.70% (1.05) | 7.94% (0.18) | 11.17% (0.52) | -14.63% (-1.55) | -3.96% (-0.59) | -5.03% (-1.62) | -5.62% (-0.91) | -7.23% (-0.65) | -2.33% (-0.18) | **3.25% (-0.37)** |
| **Frontier** | -11.37% (-2.03) | 9.60% (1.01) | 75.78% (2.65) | -2.44% (-0.31) | 2.13% (0.50) | -6.28% (-1.67) | 0.37% (0.13) | 1.03% (0.26) | 1.24% (0.25) | 4.47% (0.60) | **7.45% (0.14)** |
| Estonia | -1.93% (-0.18) | -11.44% (-0.79) | 55.63% (1.37) | 11.29% (0.40) | -3.33% (-0.20) | -17.59% (-1.95) | 0.50% (0.09) | -4.46% (-0.83) | 5.34% (0.53) | -8.70% (-1.61) | **2.53% (-0.32)** |
| Morocco | 17.57% (1.12) | 24.57% (1.92) | 9.62% (0.55) | -7.56% (-0.63) | -8.20% (-1.83) | -2.49% (-0.87) | -1.49% (-0.15) | -6.25% (-0.62) | -0.98% (-0.83) | -2.83% (-1.49) | **2.19% (-0.28)** |
| Jordan | -2.77% (-0.20) | -1.36% (-0.16) | 32.59% (1.22) | -12.14% (-1.12) | -0.37% (-0.05) | -0.28% (-0.05) | -4.27% (-1.46) | -2.94% (-1.19) | -2.37% (-0.65) | -0.49% (-0.14) | **0.56% (-0.38)** |
| **Average** | **5.77% (0.13)** | **5.15% (0.02)** | **29.71% (0.95)** | **-0.91% (-0.17)** | **-7.35% (-0.94)** | **-5.22% (-0.87)** | **-1.78% (-0.78)** | **-2.11% (-0.47)** | **-1.34% (-0.48)** | **-2.79% (-0.64)** | **1.91% (-0.33)** |

This table reports the average OOS annualized returns and Sharpe ratios of four portfolios for IS of two years and OOS of three months after transaction costs (rolling forward by one month). For example, the -4.54% annualized return of the advanced markets (2006) is calculated as the average OOS annualized return of twelve portfolios. The first portfolio's OOS return is calculated over the period 01/01/2006-31/03/2006 using as IS the period 01/01//2004-31/12/2005. The remaining eleven OOS returns are calculated by rolling forward the IS and the OOS by one month. The same logic applies for the Sharpe ratios. The last column and row present the average performance per market across all years and per year respectively.

Table 7
**Annualized Returns and Sharpe Ratios after Transaction Costs (IS 2 Years and OOS 6 Months)**

| Market | 2006 | 2007 | 2008 | 2009 | 2010 | 2011 | 2012 | 2013 | 2014 | 2015 | Average |
|---|---|---|---|---|---|---|---|---|---|---|---|
| **Advanced** | 1.05% (0.14) | -1.88% (-0.15) | 13.92% (0.57) | -2.48% (-0.22) | -9.35% (-1.49) | -5.75% (-1.16) | -2.58% (-1.14) | 0.74% (0.26) | -0.82% (-0.17) | -0.75% (-0.32) | **-0.79% (-0.37)** |
| US | 0.53% (0.06) | -6.82% (-0.46) | 26.57% (0.90) | -7.22% (-0.53) | -6.95% (-0.82) | -5.85% (-0.72) | 0.63% (0.25) | 4.77% (0.95) | 5.66% (0.75) | -6.46% (-0.95) | **0.49% (-0.06)** |
| UK | 15.09% (1.45) | -12.07% (-0.91) | 42.99% (0.98) | 1.97% (0.16) | -8.36% (-0.92) | -7.78% (-0.93) | -1.18% (-1.27) | 0.68% (0.29) | -7.72% (-1.20) | -2.01% (-0.32) | **2.16% (-0.27)** |
| Japan | -18.42% (-1.02) | -9.13% (-1.59) | 35.99% (1.11) | 0.61% (0.04) | -2.46% (-0.51) | -3.60% (-1.18) | -2.15% (-0.83) | -2.42% (-0.88) | -1.25% (-0.86) | -3.34% (-1.29) | **-0.62% (-0.70)** |
| **Emerging** | -1.43% (-0.15) | 4.61% (0.34) | 18.10% (0.57) | 0.35% (0.03) | -2.69% (-0.41) | -4.97% (-0.78) | 0.87% (0.21) | -0.65% (-0.18) | -2.32% (-0.88) | -1.87% (-0.45) | **1.00% (-0.17)** |
| Russia | 17.28% (0.68) | -2.67% (-0.20) | 34.27% (0.94) | -4.48% (-0.22) | -15.42% (-1.54) | -11.21% (-1.05) | -1.32% (-0.28) | -10.12% (-1.83) | 0.31% (0.02) | -5.83% (-0.48) | **0.08% (-0.39)** |
| China | 25.95% (1.35) | 25.27% (1.01) | 0.34% (0.01) | -5.22% (-0.33) | -3.90% (-0.48) | -5.51% (-0.76) | -0.45% (-0.91) | -4.18% (-1.71) | -5.06% (-0.78) | 0.65% (0.06) | **2.79% (-0.25)** |
| Brazil | -5.00% (-0.22) | 42.45% (0.99) | -24.49% (-0.66) | 11.68% (0.57) | -14.25% (-1.54) | -8.22% (-1.00) | -3.92% (-1.24) | -2.79% (-0.42) | -11.07% (-0.98) | 5.69% (0.40) | **-0.99% (-0.41)** |
| **Frontier** | -7.46% (-1.31) | 3.34% (0.37) | 59.03% (2.54) | -2.60% (-0.39) | 1.46% (0.34) | -6.31% (-1.78) | 1.02% (0.36) | 1.84% (0.48) | -3.06% (-0.68) | 5.85% (0.79) | **5.31% (0.07)** |
| Estonia | -2.18% (-0.18) | 0.50% (0.04) | 43.59% (1.21) | 13.47% (0.50) | -4.13% (-0.29) | -17.81% (-1.96) | 3.27% (0.58) | -5.76% (-1.20) | 0.73% (0.08) | -4.64% (-0.74) | **2.7% (-0.2)** |
| Morocco | 11.92% (0.83) | 20.50% (1.67) | -3.46% (-0.25) | -10.05% (-0.92) | -7.16% (-1.46) | -2.89% (-1.13) | -2.86% (-0.30) | -5.93% (-0.62) | -1.86% (-1.42) | -2.17% (-1.07) | **-0.4% (-0.47)** |
| Jordan | -6.88% (-0.57) | -0.36% (-0.04) | 29.07% (1.17) | -12.65% (-1.22) | 0.07% (0.01) | -3.22% (-0.61) | -4.53% (-1.75) | -2.88% (-1.14) | -2.6% (-0.76) | -0.29% (-0.09) | **-0.43% (-0.50)** |
| **Average** | **2.54% (0.09)** | **5.31% (0.09)** | **22.99% (0.76)** | **-1.39% (-0.21)** | **-6.09% (-0.76)** | **-6.93% (-1.09)** | **-1.10% (-0.53)** | **-2.23% (-0.50)** | **-2.42% (-0.57)** | **-1.27% (-0.37)** | **0.94% (-0.31)** |

This table reports the average OOS annualized returns and Sharpe ratios of two portfolios for IS of two years and OOS of six months after transaction costs (rolling forward by one month). For example, the 1.05% annualized return of the advanced markets (2006) is calculated as the average OOS annualized return of twelve portfolios. The first portfolio's OOS return is calculated over the period 01/01//2006-30/06/2006 using as IS the period 01/01//2004-31/12/2005. The remaining eleven OOS returns are calculated by rolling forward the IS and the OOS by one month. The same logic applies for the Sharpe ratios. The last column and row present the average performance per market across all years and per year respectively.

**Table 8**
**Monthly Performance Persistence for IS 2 Years (1 month rolling forward)**

| Market | 2006 | 2007 | 2008 | 2009 | 2010 | 2011 | 2012 | 2013 | 2014 | 2015 | Average |
|---|---|---|---|---|---|---|---|---|---|---|---|
| **Advanced** | 1.00 | 0.58 | 0.75 | 0.83 | 0.42 | 0.75 | 0.50 | 0.67 | 0.83 | 0.75 | **0.71** |
| US | 1.17 | 0.42 | 1.08 | 1.08 | 1.33 | 1.17 | 1.00 | 1.92 | 0.83 | 0.25 | **1.03** |
| UK | 0.92 | 0.58 | 0.50 | 1.08 | 0.42 | 0.58 | 0.08 | 0.75 | 0.42 | 1.00 | **0.63** |
| Japan | 0.17 | 0.50 | 0.58 | 0.92 | 1.58 | 0.58 | 0.25 | 0.17 | 0.08 | 0.42 | **0.53** |
| **Emerging** | 0.50 | 0.92 | 1.17 | 1.33 | 0.33 | 0.75 | 0.58 | 0.17 | 0.17 | 0.83 | **0.68** |
| Russia | 0.42 | 0.25 | 0.17 | 0.33 | 0.33 | 0.67 | 0.67 | 0.33 | 0.92 | 0.33 | **0.44** |
| China | 2.17 | 2.92 | 0.42 | 0.58 | 0.42 | 0.42 | 0.83 | 0.42 | 0.67 | 1.08 | **0.99** |
| Brazil | 0.67 | 0.58 | 0.67 | 0.75 | 0.33 | 0.42 | 0.33 | 0.58 | 0.92 | 0.50 | **0.58** |
| **Frontier** | 0.42 | 1.67 | 2.25 | 0.50 | 0.58 | 0.17 | 0.75 | 0.67 | 1.33 | 1.25 | **0.96** |
| Estonia | 0.50 | 0.42 | 0.75 | 0.58 | 0.75 | 0.17 | 0.67 | 0.42 | 1.08 | 0.25 | **0.56** |
| Morocco | 1.92 | 2.00 | 1.25 | 0.42 | 0.25 | 0.33 | 0.58 | 0.92 | 0.42 | 0.58 | **0.87** |
| Jordan | 0.42 | 0.75 | 2.08 | 0.50 | 0.83 | 0.83 | 0.25 | 0.75 | 0.58 | 0.58 | **0.76** |
| **Average** | **0.85** | **0.97** | **0.97** | **0.74** | **0.63** | **0.57** | **0.54** | **0.65** | **0.69** | **0.65** | **0.73** |

This table reports the average number of consecutive months that the monthly OOS returns of the twelve portfolio returns are above the risk-free rate. This average is calculated by generating the monthly OOS in consecutive months for each of the twelve portfolios mentioned in Table 4. For example, in advanced markets for the first portfolio, we calculate the OOS returns for 2006 (January, February, etc.). If the OOS returns over the first month are below the relevant risk-free rate, we assign a value of 0. If the OOS returns remain above the risk-free rate during the first month e.g. in January but not for February, we assign the value of 1. Otherwise, we assign a value of 2 or more. This process is repeated for the remaining eleven portfolios of the year. The analysis is done using a maximum 18 months of OOS calculations for each portfolio. The last column and row present the average monthly performance persistence per market across all years and per year respectively.

Table 9
**Quarterly Performance Persistence in Months for IS 2 2 Years (3 months rolling forward)**

| Market | 2006 | 2007 | 2008 | 2009 | 2010 | 2011 | 2012 | 2013 | 2014 | 2015 | Average |
|---|---|---|---|---|---|---|---|---|---|---|---|
| **Advanced** | 0.33 | 0.67 | 1.08 | 0.75 | 0.42 | 0.33 | 0.25 | 0.58 | 1.17 | 0.50 | **0.61** |
| US | 1.17 | 1.00 | 2.58 | 0.42 | 0.58 | 0.50 | 0.58 | 1.67 | 2.17 | 0.42 | **1.11** |
| UK | 2.17 | 0.58 | 1.17 | 0.67 | 0.33 | 0.33 | 0.00 | 0.75 | 0.42 | 0.50 | **0.69** |
| Japan | 0.33 | 0.17 | 1.33 | 0.83 | 0.67 | 0.17 | 0.17 | 0.33 | 0.08 | 0.25 | **0.43** |
| **Emerging** | 0.83 | 0.67 | 0.83 | 1.08 | 0.33 | 0.67 | 1.42 | 0.50 | 0.17 | 0.75 | **0.73** |
| Russia | 1.50 | 0.50 | 0.83 | 0.67 | 0.33 | 0.33 | 0.92 | 0.25 | 0.83 | 0.33 | **0.65** |
| China | 1.75 | 1.08 | 0.42 | 0.17 | 0.42 | 0.42 | 0.42 | 0.08 | 0.58 | 0.75 | **0.61** |
| Brazil | 1.00 | 0.58 | 0.50 | 1.08 | 0.17 | 0.25 | 0.33 | 0.58 | 0.75 | 1.00 | **0.63** |
| **Frontier** | 0.50 | 1.42 | 1.25 | 0.50 | 0.75 | 0.25 | 1.33 | 0.75 | 0.42 | 1.25 | **0.84** |
| Estonia | 0.25 | 0.67 | 1.25 | 0.75 | 0.67 | 0.08 | 0.75 | 0.25 | 1.08 | 0.25 | **0.60** |
| Morocco | 1.00 | 2.83 | 0.67 | 0.33 | 0.00 | 0.17 | 0.33 | 0.42 | 0.25 | 0.17 | **0.62** |
| Jordan | 0.42 | 0.75 | 1.17 | 0.00 | 0.42 | 0.58 | 0.42 | 0.17 | 0.67 | 0.50 | **0.51** |
| **Average** | **0.94** | **0.91** | **1.09** | **0.60** | **0.42** | **0.34** | **0.58** | **0.53** | **0.72** | **0.56** | **0.67** |

This table reports the average number of consecutive months that the quarterly OOS returns of the twelve portfolio returns are above the risk-free rate. This average is calculated by generating the quarterly OOS in consecutive quarters for each of the twelve portfolios mentioned in Table 5. For example, in advanced markets for the first portfolio, we calculate the OOS returns for 2006 (January to March, February to April, etc.). If the OOS returns over the first three months are below the relevant risk-free rate, we assign a value of 0. If the OOS returns remain above the risk-free rate only during the first 3 months of the OOS e.g. January to March but not for February to June, we assign the value of 1. Otherwise, we assign a value of 2 or more. This process is repeated for the remaining eleven portfolios of the year. The analysis is done using a maximum 18 months (6 quarters) of OOS calculations for each portfolio. The last column and row present the average performance per market across all years and per year respectively.

**Table 10**
**Semi-annual Performance Persistence in Months for IS 2 Years (6 months rolling forward)**

| Market | 2006 | 2007 | 2008 | 2009 | 2010 | 2011 | 2012 | 2013 | 2014 | 2015 | Average |
|---|---|---|---|---|---|---|---|---|---|---|---|
| **Advanced** | 0.50 | 0.42 | 1.25 | 0.50 | 0.42 | 0.08 | 0.17 | 0.75 | 0.58 | 0.33 | **0.50** |
| US | 1.00 | 0.75 | 1.67 | 0.33 | 0.42 | 0.17 | 1.17 | 2.00 | 1.83 | 0.17 | **0.95** |
| UK | 1.83 | 0.17 | 0.92 | 0.67 | 0.08 | 0.00 | 0.00 | 0.42 | 0.25 | 0.50 | **0.48** |
| Japan | 0.00 | 0.33 | 1.33 | 0.50 | 0.25 | 0.25 | 0.17 | 0.17 | 0.08 | 0.08 | **0.32** |
| **Emerging** | 1.25 | 0.67 | 1.42 | 1.00 | 0.50 | 0.42 | 1.25 | 0.42 | 0.67 | 0.25 | **0.78** |
| Russia | 1.25 | 0.42 | 1.17 | 0.50 | 0.08 | 0.33 | 0.67 | 0.00 | 0.67 | 0.17 | **0.53** |
| China | 1.58 | 0.83 | 1.17 | 0.42 | 0.50 | 0.08 | 0.00 | 0.00 | 0.08 | 0.83 | **0.55** |
| Brazil | 1.33 | 1.83 | 0.67 | 1.00 | 0.08 | 0.00 | 0.08 | 0.25 | 0.25 | 1.25 | **0.68** |
| **Frontier** | 0.67 | 0.75 | 1.50 | 0.33 | 0.50 | 0.08 | 1.08 | 1.33 | 0.75 | 1.00 | **0.80** |
| Estonia | 0.50 | 0.75 | 2.08 | 1.25 | 0.50 | 0.00 | 0.67 | 0.00 | 0.67 | 0.42 | **0.68** |
| Morocco | 1.58 | 1.83 | 0.42 | 0.00 | 0.00 | 0.08 | 0.50 | 0.17 | 0.00 | 0.08 | **0.47** |
| Jordan | 0.33 | 1.00 | 1.83 | 0.00 | 0.67 | 0.25 | 0.00 | 0.00 | 0.75 | 1.08 | **0.59** |
| **Average** | **0.99** | **0.81** | **1.28** | **0.54** | **0.33** | **0.15** | **0.48** | **0.46** | **0.55** | **0.51** | **0.61** |

This table reports the average number of consecutive months that the semi-annual OOS returns of the twelve portfolio returns are above the risk-free rate. This average is calculated by generating the semi-annual OOS in consecutive quarters for each of the twelve portfolios mentioned in Table 6. For example, in advanced markets for the first portfolio, we calculate the OOS returns for 2006 (January to June, February to July, etc.). If the OOS returns over the first six months are below the relevant risk-free rate, we assign a value of 0. If the OOS returns remain above the risk-free rate only during the first 6 months of the OOS e.g. January to June but not for July to December, we assign the value of 1. Otherwise, we assign a value of 2 or more. This process is repeated for the remaining eleven portfolios of the year. The analysis is done using a maximum 18 months (3 semesters) of OOS calculations for each portfolio. The last column and row present the average performance per market across all years and per year respectively.

**Table 11**
**Annualized Returns based on the cross-validated surviving rules (IS of 2 Years and OOS 1 Month)**

| Market | 2006 | 2007 | 2008 | 2009 | 2010 | 2011 | 2012 | 2013 | 2014 | 2015 | Average |
|---|---|---|---|---|---|---|---|---|---|---|---|
| **Advanced** | 27.05% (0.02%) | 23.08% (0.38%) | 103.68% (0.09%) | 93.3% (1%) | 45.17% (0.73%) | 46.44% (3.19%) | 36.01% (0.03%) | 17.92% (3.22%) | 21.53% (7.76%) | 27.87% (0.06%) | **44.21% (1.65%)** |
| US | 8.19% (0%) | 18.04% (0.01%) | 87.67% (0.04%) | 84.56% (0.33%) | 50.42% (0.38%) | 34.72% (4.82%) | 24.13% (0.05%) | 26.95% (7.41%) | 21.08% (17.41%) | 21.33% (8.32%) | **37.71% (3.88%)** |
| UK | 23.8% (9.29%) | 27.1% (6.91%) | 89.14% (0.08%) | 85.49% (6.76%) | 56.82% (7.79%) | 50.13% (5.17%) | 27.01% (0.01%) | 20.6% (0.03%) | 40.89% (4.1%) | 36.97% (0.07%) | **45.8% (4.02%)** |
| Japan | 29.93% (0.02%) | 6.87% (0.02%) | 59.93% (0.02%) | 66.47% (1.3%) | 39.6% (0.61%) | 32.77% (0.14%) | 36.79% (0.02%) | 29.92% (0.03%) | 6.11% (0.01%) | 9.93% (0.01%) | **31.83% (0.22%)** |
| **Emerging** | 58.41% (0.58%) | 70.48% (0.57%) | 163.42% (0.54%) | 116.68% (1.7%) | 51.44% (2.99%) | 53.04% (1.74%) | 39.71% (0.08%) | 27.96% (0.09%) | 30.23% (0.05%) | 47.04% (0.04%) | **65.84% (0.84%)** |
| Russia | 86.02% (4.44%) | 27.64% (2.67%) | 140.78% (0.85%) | 183.65% (8.59%) | 67.8% (6.09%) | 74.31% (6.62%) | 75.11% (0.06%) | 44.42% (0.11%) | 109.66% (0.13%) | 91.44% (0.23%) | **90.08% (2.98%)** |
| China | 99.78% (2.08%) | 99.23% (18.07%) | 170% (1.23%) | 104.07% (0.35%) | 55.13% (0.18%) | 49.38% (1.62%) | 43.64% (0.01%) | 27.92% (0.03%) | 37.37% (0.07%) | 78.59% (0.21%) | **76.51% (2.38%)** |
| Brazil | 106.43% (6.81%) | 103.96% (3.32%) | 172.44% (1.62%) | 122.07% (6.6%) | 65.04% (7.76%) | 39.45% (3.15%) | 36.28% (0.01%) | 51.21% (0.08%) | 59.73% (0.31%) | 79.44% (0.08%) | **83.6% (2.97%)** |
| **Frontier** | 34.04% (5.02%) | 45.26% (0.65%) | 131.42% (2.91%) | 90.36% (14.16%) | 40.02% (10.53%) | 30.28% (2.11%) | 26.5% (0.12%) | 25.28% (0.18%) | 26.89% (2.37%) | 35.09% (0.71%) | **48.51% (3.88%)** |
| Estonia | 42.02% (7.72%) | 64.11% (0.1%) | 246.5% (2.06%) | 119.29% (3.52%) | 141.52% (3.22%) | 52.77% (1.96%) | 43.17% (0.04%) | 31.11% (1.68%) | 51.85% (2.55%) | 29.85% (0.03%) | **82.22% (2.29%)** |
| Morocco | 97.25% (4.06%) | 42.05% (18.14%) | 77.99% (9.32%) | 51.52% (1.75%) | 37.84% (0.11%) | 16% (0.02%) | 36.57% (0.07%) | 40.51% (0.29%) | 10.15% (0.01%) | 16.5% (0.03%) | **42.64% (3.38%)** |
| Jordan | 89.39% (7.42%) | 45.23% (0.46%) | 103.32% (0.93%) | 52.78% (1.82%) | 35.76% (0.59%) | 37.43% (0.24%) | 27.1% (0.05%) | 14.9% (0.01%) | 18.51% (0.03%) | 22.1% (0.03%) | **44.65% (1.16%)** |
| **Average** | **58.52% (3.95%)** | **47.75% (4.27%)** | **128.86% (1.64%)** | **97.52% (3.99%)** | **57.21% (3.41%)** | **43.06% (2.56%)** | **37.67% (0.05%)** | **29.89% (1.1%)** | **36.17% (2.9%)** | **41.35% (0.82%)** | **57.80% (2.47%)** |

This table reports the average OOS annualized returns of twelve cross-validated portfolios for IS of two years and OOS of one month after transaction costs (rolling forward by one month). In parentheses we report the average percentage of cross-validated rules from the genuine significant rules as identified in section 5.1. For example, Table 2 reports that 2628 rules (0.124*21195) survive on average in the case of frontier markets (2006). This table estimates that out of those rules, 132 (0.0502*2628) survive both in the IS and OOS and generate an average OOS annualized return of 34.04%.

**Table 12**
**Annualized Returns based on the cross-validated surviving rules (IS of 2 Years and OOS 3 Months)**

| Market | 2006 | 2007 | 2008 | 2009 | 2010 | 2011 | 2012 | 2013 | 2014 | 2015 | Average |
|---|---|---|---|---|---|---|---|---|---|---|---|
| **Advanced** | 12.96% (0.02%) | 15.73% (0.04%) | 64.58% (0.3%) | 36.66% (0.85%) | 21.51% (0.6%) | 22.21% (0.09%) | 20.42% (0.04%) | 12.8% (3.17%) | 9.11% (7.13%) | 12.15% (0.07%) | **22.81% (1.23%)** |
| US | 7.35% (0%) | 6.38% (0%) | 65.08% (0.12%) | 32.66% (0.25%) | 22.13% (0.36%) | 14.94% (4.77%) | 11.58% (0.07%) | 17.34% (5.61%) | 9.81% (25.83%) | 10.4% (9.17%) | **19.77% (4.62%)** |
| UK | 19% (15.16%) | 22.96% (6.83%) | 69.49% (0.18%) | 46.82% (5.78%) | 26.64% (7.74%) | 18.44% (5.03%) | 9.38% (0.01%) | 16.23% (0.02%) | 14.79% (0.22%) | 20.89% (0.08%) | **26.46% (4.11%)** |
| Japan | 10.12% (0.76%) | 3.73% (0.02%) | 41.04% (0.28%) | 43.49% (1.04%) | 22.62% (0.62%) | 12.87% (0.11%) | 19.49% (0.03%) | 11.58% (0.03%) | 3.39% (0.01%) | 5.76% (0.01%) | **17.41% (0.29%)** |
| **Emerging** | 31.64% (0.43%) | 38.19% (0.57%) | 75.18% (0.63%) | 50.2% (1.66%) | 23.89% (1.71%) | 34.06% (4.92%) | 23.63% (0.07%) | 18.35% (0.1%) | 15.51% (0.04%) | 20.47% (0.06%) | **33.11% (1.02%)** |
| Russia | 44.72% (9.77%) | 14.1% (6.37%) | 53.41% (1.04%) | 104.76% (12.61%) | 31.66% (5.06%) | 39.9% (3.35%) | 32.95% (0.07%) | 25.2% (0.1%) | 60.47% (0.13%) | 42.05% (0.2%) | **44.92% (3.87%)** |
| China | 54.69% (2.01%) | 74.44% (15.23%) | 75.41% (0.86%) | 46.09% (0.36%) | 22.33% (0.19%) | 25.56% (3.19%) | 19.85% (0.01%) | 14.43% (0.03%) | 17.94% (0.05%) | 49.18% (0.27%) | **39.99% (2.22%)** |
| Brazil | 57.35% (10.46%) | 68.62% (3.21%) | 87.29% (2.16%) | 57.1% (8.46%) | 23.59% (7.2%) | 13.57% (3.13%) | 16.7% (0.01%) | 29.39% (0.08%) | 28.77% (0.37%) | 43.54% (0.09%) | **42.59% (3.52%)** |
| **Frontier** | 21.22% (1.96%) | 28.95% (0.56%) | 108.66% (3.44%) | 33.05% (14.01%) | 21.13% (11.5%) | 15.07% (2.29%) | 16.44% (0.14%) | 15.11% (0.16%) | 18.46% (0.73%) | 20.69% (0.73%) | **29.88% (3.55%)** |
| Estonia | 30.96% (2.65%) | 30.7% (0.04%) | 99.4% (2.55%) | 63.69% (3.78%) | 49.29% (2.38%) | 23.41% (0.3%) | 20.84% (0.05%) | 14.3% (0.12%) | 27.51% (2.27%) | 8% (0.04%) | **36.81% (1.42%)** |
| Morocco | 50% (2.44%) | 29.76% (20.98%) | 52.44% (6.93%) | 18.75% (1.65%) | 17.55% (0.1%) | 6.13% (0.02%) | 15.83% (0.08%) | 17.1% (0.24%) | 7.59% (0.01%) | 3.75% (0.01%) | **21.89% (3.25%)** |
| Jordan | 41.88% (8.19%) | 21.65% (0.28%) | 65.41% (0.91%) | 16.62% (1.38%) | 20.99% (0.56%) | 17.82% (0.26%) | 11.97% (0.06%) | 11.41% (0.01%) | 7.3% (0.03%) | 7.93% (0.03%) | **22.3% (1.17%)** |
| **Average** | **31.82% (4.49%)** | **29.6% (4.51%)** | **71.45% (1.62%)** | **45.82% (4.32%)** | **25.28% (3.17%)** | **20.33% (2.29%)** | **18.26% (0.05%)** | **16.94% (0.81%)** | **18.39% (3.07%)** | **20.4% (0.9%)** | **29.83% (2.52%)** |

This table reports the average OOS annualized returns of twelve cross-validated portfolios for IS of two years and OOS of one month after transaction costs (rolling forward by one month). In parentheses we report the average percentage of cross-validated rules from the total pool of rules. For example, Table 2 reports that 2628 rules (0.124*21195) survive on average in the case of frontier markets (2006). This table estimates that out of those rules, 51 (0.0196*2628) survive both in the IS and OOS and generate an average OOS annualized return of 21.22%.

**Table 13**
**Annualized Returns based on the cross-validated surviving rules (IS of 2 Years and OOS 6 Months)**

| Market | 2006 | 2007 | 2008 | 2009 | 2010 | 2011 | 2012 | 2013 | 2014 | 2015 | Average |
|---|---|---|---|---|---|---|---|---|---|---|---|
| **Advanced** | 13.22% (0.02%) | 8.9% (0.03%) | 45.74% (0.08%) | 22.7% (0.73%) | 12.31% (0.45%) | 12.04% (0.07%) | 10.16% (0.02%) | 8.82% (3.21%) | 5.13% (10.2%) | 10.06% (0.07%) | 14.91% (1.49%) |
| US | 4.97% (0%) | 7.09% (0.01%) | 50.45% (0.03%) | 20.32% (0.18%) | 13.2% (0.25%) | 11.41% (0.07%) | 9.18% (0.07%) | 13.54% (7.34%) | 7.65% (27.92%) | 9.06% (4.63%) | 14.69% (4.05%) |
| UK | 16.89% (17.58%) | 15.15% (3.46%) | 57.18% (0.19%) | 30.06% (4.59%) | 17.02% (7.38%) | 11.72% (0.21%) | 7.73% (0.01%) | 12.62% (0.03%) | 10.12% (0.1%) | 15.6% (0.07%) | 19.41% (3.36%) |
| Japan | 1.65% (0.02%) | 1.17% (0.01%) | 40.34% (0.17%) | 26.61% (1.12%) | 12.19% (0.48%) | 7.81% (0.12%) | 10.42% (0.04%) | 5.66% (0.03%) | 2.52% (0.02%) | 5.63% (0.02%) | 11.4% (0.2%) |
| **Emerging** | 19.85% (0.46%) | 31.44% (0.77%) | 68.91% (0.33%) | 28.07% (1.49%) | 14.62% (3.44%) | 26.61% (1.8%) | 17.52% (0.08%) | 11.14% (0.12%) | 11.06% (0.03%) | 16.32% (0.05%) | 24.55% (0.86%) |
| Russia | 23.72% (9.25%) | 10% (5.9%) | 60.63% (1.2%) | 59.59% (10.61%) | 21.97% (4.92%) | 21.17% (3.3%) | 18.7% (0.06%) | 14.59% (0.08%) | 42.1% (0.11%) | 27.76% (0.22%) | 30.02% (3.56%) |
| China | 36.6% (2.42%) | 56.76% (14.35%) | 54.33% (0.7%) | 24.73% (0.38%) | 12.09% (1.74%) | 14.03% (0.08%) | 8.3% (0.01%) | 9.2% (0.02%) | 17.33% (0.06%) | 30.88% (0.28%) | 26.42% (2%) |
| Brazil | 26.05% (6.44%) | 50.74% (5.66%) | 57.57% (2.61%) | 33.43% (10.36%) | 13.81% (5.86%) | 12.13% (0.06%) | 10.35% (0.01%) | 20.98% (0.13%) | 18.28% (0.3%) | 49.05% (0.1%) | 29.24% (3.15%) |
| **Frontier** | 16.76% (1.61%) | 18.92% (0.52%) | 86.73% (2.72%) | 19.73% (13.14%) | 12.41% (13.1%) | 9.47% (0.73%) | 11.17% (0.14%) | 13.55% (0.23%) | 10.2% (0.46%) | 16.49% (0.54%) | 21.54% (3.32%) |
| Estonia | 23.28% (1.38%) | 25.66% (0.05%) | 78.2% (1.96%) | 43.24% (4.44%) | 23.12% (2.67%) | 12.24% (0.25%) | 14.04% (0.07%) | 6.54% (0.11%) | 17.91% (1.56%) | 7.16% (0.04%) | 25.14% (1.25%) |
| Morocco | 35.66% (2.28%) | 22.89% (25.24%) | 34.85% (5.99%) | 9.48% (1.34%) | 12.06% (0.07%) | 4.82% (0.02%) | 9.27% (0.1%) | 11.46% (0.07%) | 1.84% (0.01%) | 3.39% (0.02%) | 14.57% (3.51%) |
| Jordan | 26.1% (7.34%) | 14.5% (0.24%) | 54% (1.05%) | 10.79% (1.03%) | 13.75% (0.58%) | 10.64% (0.23%) | 5.9% (0.04%) | 5.08% (0.01%) | 6.5% (0.02%) | 5.79% (0.03%) | 15.31% (1.06%) |
| **Average** | 20.4% (4.07%) | 21.94% (4.69%) | 57.41% (1.42%) | 27.4% (4.12%) | 14.88% (3.41%) | 12.84% (0.58%) | 11.06% (0.06%) | 11.1% (0.95%) | 12.55% (3.4%) | 16.43% (0.51%) | 20.6% (2.32%) |

This table reports the average OOS annualized returns of twelve cross-validated portfolios for IS of two years and OOS of one month after transaction costs (rolling forward by one month). In parentheses we report the average percentage of cross-validated rules from the total pool of rules. For example, Table 2 reports that 2628 rules (0.124*21195) survive on average in the case of frontier markets (2006). This table estimates that out of those rules, 42 (0.0161*2628) survive both in the IS and OOS and generate an average OOS annualized return of 16.76%.

**Table 14**
**Financial Stress Performance**

| Market | Period | Financial Stress | 2006 | 2007 | 2008 | 2009 | 2010 | 2011 | 2012 | 2013 | 2014 | 2015 | Average |
|---|---|---|---|---|---|---|---|---|---|---|---|---|---|
| US | IS of 2 Years - OOS 1 Month | High | - | -9.96% (-0.59) | 22.72% (0.77) | -3.64% (-0.19) | -18.27% (-2.18) | -0.57% (-0.17) | -5.08% (-2.16) | - | - | -3.86% (-0.75) | -2.67% (-0.75) |
| US | | Low | 1.81% (0.2) | -3.36% (-0.32) | - | - | -0.28% (-0.02) | -2.42% (-0.39) | 0.53% (0.18) | 7.02% (1.31) | 5.73% (0.77) | -5.41% (-0.71) | 0.45% (0.13) |
| Other Advanced | | High | - | -0.72% (-0.06) | 10.63% (0.41) | 4.14% (0.25) | -16.24% (-2.88) | -2.15% (-0.55) | -4.86% (-1.81) | - | - | -0.37% (-0.07) | -1.37% (-0.59) |
| Other Advanced | | Low | -3.66% (-0.54) | -6.02% (-0.57) | - | - | -33.23% (-3.59) | 0.61% (0.16) | 0.11% (0.05) | 0.35% (0.13) | -2.07% (-0.42) | -1.53% (-0.85) | -5.68% (-0.7) |
| Emerging | | High | - | -13.42% (-4.45) | 43.52% (1.1) | 1.78% (0.11) | -20.32% (-3.13) | -4.89% (-0.69) | 3% (0.74) | 26.12% (4.45) | -4.83% (-4.01) | -1.48% (-0.37) | 3.28% (-0.69) |
| Emerging | | Low | 1.75% (0.15) | -0.82% (-0.06) | - | 7.8% (0.7) | -4.37% (-0.66) | -1.87% (-0.36) | -3.65% (-0.94) | -3.94% (-1.45) | -3.61% (-1.21) | 0.48% (0.29) | -0.91% (-0.39) |

This table reports the average OOS annualized returns and Sharpe ratios of the portfolios generated in Section 5.1. High and low correspond to high and low financial stress conditions as reported by the OFR stress indices. – indicates that for this year and market there were no periods with high (or low) financial stress.

**Table A.1**
**Annualized mean excess returns for quartiles of different combination of Sharpe ratio levels.**

| Outperforming SR | Quartile | Underperforming SR | | | | | |
|---|---|---|---|---|---|---|---|
| | | -2 | | -3 | | -4 | |
| | | Outperforming | Underperforming | Outperforming | Underperforming | Outperforming | Underperforming |
| 2 | 1st | 6.50 | -8.93 | 6.80 | -16.28 | 6.60 | -24.69 |
| | 2nd | 16.28 | -16.08 | 16.92 | -23.84 | 16.51 | -32.74 |
| | 3rd | 19.14 | -22.55 | 19.83 | -30.71 | 19.25 | -39.62 |
| 3 | 1st | 11.51 | -8.69 | 11.37 | -16.83 | 11.43 | -24.64 |
| | 2nd | 25.14 | -15.86 | 24.63 | -24.52 | 24.81 | -32.82 |
| | 3rd | 27.94 | -22.53 | 27.38 | -31.05 | 27.54 | -39.55 |
| 4 | 1st | 16.44 | -8.88 | 16.32 | -16.66 | 16.64 | -24.19 |
| | 2nd | 33.56 | -16.00 | 33.43 | -24.34 | 34.22 | -32.23 |
| | 3rd | 36.19 | -22.47 | 36.08 | -30.97 | 36.90 | -39.13 |

This table reports the quartiles of the distribution of the annualized mean excess return (in percentages) induced by positive and negative Sharpe ratio pairs applied in the Monte Carlo simulations for the out- and under-performing strategies. The pairs are created with the annualized Sharpe ratio for out- and under-performing rules set to 2, 3, 4 and -2, -3, -4 respectively. The quantities presented correspond to the average values over 1000 Monte Carlo simulations. The proportion of rules that are neutrally performing ($\pi_0$), outperforming ($\pi_A^+$) and underperforming ($\pi_A^-$) and are set to 50%, 20% and 30% respectively.

**Table.A2**
**Estimation of neutral, positive, and negative proportions by the DFDR$^{+/-}$ procedure versus the actual ones.**

| Outperforming SR | Proportion | Underperforming SR | | |
|---|---|---|---|---|
| | | -2 | -3 | -4 |
| 2 | $\pi_0 = 50\%$ | 73.08 | 62.93 | 60.97 |
| | $\pi_A^+ = 20\%$ | 9.39 | 11.91 | 11.17 |
| | $\pi_A^- = 30\%$ | 17.53 | 25.17 | 27.86 |
| 3 | $\pi_0 = 50\%$ | 66.43 | 57.09 | 53.44 |
| | $\pi_A^+ = 20\%$ | 14.35 | 15.23 | 16.09 |
| | $\pi_A^- = 30\%$ | 19.22 | 27.68 | 30.47 |
| 4 | $\pi_0 = 50\%$ | 64.35 | 53.55 | 50.21 |
| | $\pi_A^+ = 20\%$ | 16.28 | 18.08 | 18.89 |
| | $\pi_A^- = 30\%$ | 19.37 | 28.36 | 30.91 |

The quantities presented correspond to the average values estimated over 1000 Monte Carlo simulations. The proportion of rules that are neutrally performing ($\pi_0$), outperforming ($\pi_A^+$) and underperforming ($\pi_A^-$) are set to 50%, 20% and 30% respectively. The table provides the estimates when the annualized Sharpe ratio for out- and under-performing rules is set to 2, 3, 4 and -2, -3, -4 respectively.

**Table A.3**
**True false discovery rate, accuracy and the positive-performing portfolio size through different methods.**

| Outperforming SR | Portfolio Type | Underperforming SR | | | | | | | | |
|---|---|---|---|---|---|---|---|---|---|---|
| | | -2 | | | -3 | | | -4 | | |
| | | FDR$^+$ | Power | Portfolio size | FDR$^+$ | Power | Portfolio size | FDR$^+$ | Power | Portfolio size |
| 2 | 10%-DFDR$^+$ | 12.92 | 39.46 | 1995.71 | 12.67 | 42.01 | 2118.86 | 12.05 | 40.31 | 2014.43 |
| | 10%-FDR | 16.36 | 11.75 | 896.75 | 14.14 | 11.98 | 941.47 | 14.61 | 11.91 | 883.09 |
| | 20%-DFDR$^+$ | 13.79 | 39.92 | 2110.78 | 15.03 | 43.3 | 2369.41 | 14.65 | 41.66 | 2281.92 |
| | 20%-FDR | 16.12 | 16.67 | 896.75 | 13.95 | 17.69 | 941.47 | 14.63 | 16.15 | 883.09 |
| | 5%-RW | 0.86 | 0.01 | 0.53 | 0.90 | 0.01 | 0.52 | 0.71 | 0.01 | 0.48 |
| | 20%-RW | 8.42 | 0.05 | 3.03 | 8.28 | 0.06 | 3.33 | 7.11 | 0.05 | 2.90 |
| 3 | 10%-DFDR$^+$ | 8.44 | 64.74 | 3076.41 | 8.00 | 64.74 | 3048.45 | 8.78 | 62.89 | 3039.27 |
| | 10%-FDR | 8.29 | 29.61 | 1856.71 | 7.3 | 30.27 | 1906.89 | 7.59 | 31.46 | 1945.02 |
| | 20%-DFDR$^+$ | 9.54 | 65.29 | 3212.88 | 10.59 | 66.30 | 3321.58 | 11.55 | 64.62 | 3343.04 |
| | 20%-FDR | 8.89 | 38.26 | 1856.71 | 7.9 | 39.86 | 1906.89 | 8.25 | 40.11 | 1945.02 |
| | 5%-RW | 0.31 | 0.01 | 0.52 | 0.04 | 0.01 | 0.60 | 0.08 | 0.01 | 0.57 |
| | 20%-RW | 3.21 | 0.07 | 3.47 | 2.30 | 0.07 | 3.41 | 2.97 | 0.07 | 3.49 |
| 4 | 10%-DFDR$^+$ | 6.45 | 82.39 | 3790.07 | 6.62 | 83.57 | 3879.56 | 7.83 | 83.01 | 3945.84 |
| | 10%-FDR | 4.85 | 56.8 | 3187.43 | 4.56 | 54.55 | 3012.02 | 4.50 | 55.99 | 3138.75 |
| | 20%-DFDR$^+$ | 7.80 | 83.35 | 3948.29 | 9.53 | 85.49 | 4194.29 | 10.5 | 84.89 | 4244.14 |
| | 20%-FDR | 6.63 | 68.06 | 3187.43 | 5.6 | 65.59 | 3012.02 | 5.96 | 67.98 | 3138.75 |
| | 5%-RW | 0.00 | 0.02 | 0.65 | 0.01 | 0.02 | 0.73 | 0.00 | 0.02 | 0.68 |
| | 20%-RW | 0.53 | 0.09 | 3.87 | 0.31 | 0.10 | 4.38 | 0.34 | 0.09 | 3.73 |

This table reports the FDR$^+$ and accuracy in percentages and the portfolio size (out of 21195). Accuracy is estimated by the ratio of actual outperformers discovered by the underlying procedure. We consider confidence target levels of 10% and 20% for the DFDR$^+$ and benchmark it against the FDR procedure in Storey *et al.* (2004) for the same target levels. The second benchmark is the FWER procedure by RW tested at the levels 5% and 20%. The quantities refer to average values over 1000 Monte Carlo simulations for different combinations pairs, when annualized Sharpe ratio for out- and under-performing rules is set to 2, 3, 4 and -2, -3, -4 respectively.

**Table B.1**
**Percentage and standard deviation of the DFDR$^{+/-}$ procedure survivors (IS 1 year).**

| Market | 2006 | 2007 | 2008 | 2009 | 2010 | 2011 | 2012 | 2013 | 2014 | 2015 | Average |
|---|---|---|---|---|---|---|---|---|---|---|---|
| **Advanced** | 1.37 (2.82) | 0.20 (0.20) | 1.34 (3.22) | 5.44 (4.12) | 4.32 (7.72) | 2.10 (5.26) | 0.42 (0.20) | 8.93 (11.28) | 7.72 (12.61) | 0.45 (0.12) | **3.23 (4.76)** |
| US | 0.02 (0.02) | 0.11 (0.12) | 0.57 (1.27) | 2.53 (2.68) | 6.02 (9.25) | 2.15 (5.61) | 2.06 (5.34) | 16.75 (15.89) | 16.24 (16.68) | 0.32 (0.13) | **4.68 (5.70)** |
| UK | 11.52 (6.91) | 14.52 (11.99) | 2.75 (6.43) | 39.02 (9.20) | 13.40 (14.60) | 0.57 (0.23) | 0.24 (0.07) | 2.29 (6.25) | 3.17 (8.59) | 0.37 (0.11) | **8.78 (6.44)** |
| Japan | 13.19 (13.86) | 1.26 (2.59) | 2.70 (4.57) | 7.01 (2.60) | 4.90 (7.97) | 0.66 (0.22) | 0.49 (0.20) | 2.27 (5.30) | 0.23 (0.10) | 0.44 (0.15) | **3.32 (3.76)** |
| **Emerging** | 4.87 (9.35) | 1.94 (2.38) | 1.01 (1.33) | 6.74 (5.01) | 10.21 (11.65) | 0.48 (0.30) | 0.61 (0.19) | 0.37 (0.21) | 0.32 (0.17) | 0.53 (0.67) | **2.71 (3.12)** |
| Russia | 31.04 (19.00) | 1.92 (3.65) | 2.81 (6.59) | 33.80 (12.19) | 11.24 (12.74) | 2.65 (5.77) | 0.75 (0.45) | 0.94 (0.63) | 0.96 (0.38) | 1.39 (0.77) | **8.75 (6.22)** |
| China | 7.34 (8.51) | 27.46 (23.36) | 0.96 (0.61) | 2.90 (5.44) | 5.68 (9.01) | 0.46 (0.70) | 0.37 (0.08) | 0.91 (0.47) | 0.31 (0.25) | 5.60 (9.06) | **5.20 (5.75)** |
| Brazil | 33.66 (30.99) | 8.85 (12.55) | 16.00 (15.17) | 34.05 (9.11) | 10.67 (11.65) | 0.26 (0.20) | 0.40 (0.40) | 0.97 (0.53) | 1.44 (0.61) | 1.24 (1.46) | **10.75 (8.27)** |
| **Frontier** | 9.18 (14.08) | 6.12 (8.23) | 3.58 (9.53) | 38.73 (14.25) | 7.03 (8.80) | 1.64 (0.80) | 0.51 (0.30) | 3.05 (5.17) | 8.17 (9.53) | 2.80 (1.58) | **8.08 (7.23)** |
| Estonia | 0.66 (1.28) | 0.83 (0.50) | 6.04 (9.17) | 7.08 (5.44) | 10.04 (10.78) | 1.09 (0.98) | 0.47 (0.30) | 25.02 (8.96) | 1.76 (2.35) | 1.50 (2.86) | **5.45 (4.26)** |
| Morocco | 19.28 (22.47) | 4.67 (4.39) | 12.35 (10.54) | 2.02 (1.55) | 0.31 (0.23) | 0.18 (0.06) | 1.84 (2.17) | 0.22 (0.06) | 0.42 (0.16) | 0.59 (0.82) | **4.19 (4.24)** |
| Jordan | 0.89 (1.19) | 1.25 (1.03) | 2.38 (2.98) | 4.38 (2.67) | 0.39 (0.11) | 0.82 (0.96) | 0.35 (0.13) | 0.26 (0.21) | 0.27 (0.10) | 0.38 (0.21) | **1.14 (0.96)** |
| **Average** | **11.08 (10.87)** | **5.76 (5.92)** | **4.37 (5.95)** | **15.31 (6.19)** | **7.02 (8.71)** | **1.09 (1.76)** | **0.71 (0.82)** | **5.17 (4.58)** | **3.42 (4.29)** | **1.30 (1.49)** | **5.52 (5.06)** |

This table reports the percentage and standard deviations of the survivor rules adjusted based on the number of the total number rules. For example, in 2006 for the advanced market the surviving rules are 290 (0.0137*21195) and their standard deviation is 598 (0.0282*21195). The average is estimated from the twelve portfolios whose OOS is in 2006. The first portfolio's IS runs from 01/01/2004-31/12/2005 and the remaining eleven are calculated by rolling forward the IS by one month.

**Table B.2**
**Annualized Returns and Sharpe Ratios after Transaction Costs (IS 1 Year)**

| Market | 2006 | 2007 | 2008 | 2009 | 2010 | 2011 | 2012 | 2013 | 2014 | 2015 | Average |
|---|---|---|---|---|---|---|---|---|---|---|---|
| **Advanced** | 16.19% (2.04) | 16.91% (1.97) | 23.67% (1.92) | 50.74% (2.18) | 16.38% (2.36) | 14.66% (2.13) | 12.49% (2.27) | 15.27% (2.28) | 9.74% (2.31) | 7.01% (3.14) | **18.31% (2.26)** |
| US | 11.78% (1.35) | 14.38% (1.56) | 23.21% (1.43) | 48.31% (1.65) | 16.47% (1.86) | 15.01% (2.21) | 12.84% (2.20) | 14.94% (2.03) | 12.91% (1.84) | 5.48% (2.33) | **17.53% (1.85)** |
| UK | 10.71% (1.31) | 15.37% (1.52) | 18.59% (0.95) | 49.79% (1.41) | 22.49% (1.86) | 19.64% (1.97) | 7.26% (2.42) | 12.81% (2.04) | 11.88% (2.58) | 11.17% (2.85) | **17.97% (1.89)** |
| Japan | 23.33% (1.36) | 13.65% (0.93) | 13.77% (1.04) | 34.03% (1.39) | 12.38% (1.31) | 15.59% (1.68) | 7.84% (2.21) | 16.86% (1.90) | 5.81% (2.35) | 8.31% (2.19) | **15.16% (1.64)** |
| **Emerging** | 29.38% (2.6) | 33.62% (2.51) | 41.64% (2.19) | 69.6% (2.51) | 20.41% (2.54) | 18.18% (2.35) | 21.84% (2.49) | 15.57% (2.59) | 9.46% (2.46) | 13.92% (2.48) | **27.36% (2.47)** |
| Russia | 39.42% (1.26) | 26.95% (1.19) | 28.83% (0.95) | 99.79% (1.78) | 27.57% (1.43) | 28.66% (2.12) | 31.19% (2.26) | 23.39% (2.02) | 17.13% (2.23) | 47.18% (2.18) | **37.01% (1.74)** |
| China | 34.17% (2.22) | 41.65% (2.32) | 55.04% (2.10) | 69.05% (1.90) | 19.00% (1.86) | 10.95% (1.98) | 13.55% (2.70) | 19.17% (2.07) | 18.40% (2.34) | 23.90% (1.97) | **30.49% (2.14)** |
| Brazil | 28.63% (1.42) | 31.85% (1.22) | 40.21% (1.22) | 68.74% (1.51) | 24.13% (1.88) | 12.54% (1.76) | 29.99% (2.48) | 17.52% (2.17) | 21.09% (2.48) | 32.27% (1.69) | **30.7% (1.78)** |
| **Frontier** | 24.2% (2.69) | 25.08% (2.63) | 28.36% (2.27) | 55.37% (2.87) | 21.1% (3.21) | 16.34% (2.53) | 13.5% (2.30) | 12.6% (2.82) | 10.67% (2.85) | 17.95% (2.46) | **22.52% (2.66)** |
| Estonia | 18.92% (1.87) | 38.79% (2.27) | 44.25% (2.25) | 75.54% (2.36) | 38.34% (2.14) | 33.91% (2.17) | 15.69% (2.35) | 19.76% (1.93) | 11.87% (1.63) | 15.01% (1.89) | **31.21% (2.09)** |
| Morocco | 36.42% (3.04) | 39.55% (2.48) | 32.95% (2.13) | 37.99% (2.16) | 12.64% (2.35) | 8.85% (2.31) | 18.02% (1.64) | 6.82% (2.35) | 8.32% (2.04) | 9.68% (1.31) | **21.12% (2.18)** |
| Jordan | 41.04% (2.21) | 29.06% (1.92) | 32.41% (2.15) | 46.15% (2.02) | 16.23% (2.40) | 17.99% (2.28) | 12.73% (2.21) | 8.33% (1.99) | 10.38% (1.84) | 5.46% (2.54) | **21.98% (2.16)** |
| **Average** | **26.18% (1.95)** | **27.24% (1.88)** | **31.91% (1.72)** | **58.76% (1.98)** | **20.60% (2.10)** | **17.69% (2.12)** | **16.41% (2.30)** | **15.25% (2.18)** | **12.31% (2.25)** | **16.45% (2.25)** | **24.28% (2.07)** |

This table reports the average IS annualized returns and Sharpe ratios of twelve portfolios for one year of IS after transaction costs (rolling forward by one month). For example, the 16.19% annualized return of the advanced markets (2006) is calculated as the average IS annualized return of twelve portfolios. The first portfolio's IS return is calculated over the period of 01/01/2005-31/12/2005. The remaining eleven are calculated by rolling forward the IS by one month. The same logic applies for the Sharpe ratios. The last column and row present the average performance per market across all years and per year respectively.

Table B.3
**Percentage decomposition of RSI rules among the DFDR$^{+/-}$ procedure survivors (IS 2 Years)**

| Average of RSI % | 2006 | 2007 | 2008 | 2009 | 2010 | 2011 | 2012 | 2013 | 2014 | 2015 |
|---|---|---|---|---|---|---|---|---|---|---|
| **Advanced** | 0.12% | 0.00% | 0.00% | 0.53% | 1.63% | 1.27% | 6.32% | 0.43% | 0.15% | 39.29% |
| US | 0.00% | 0.00% | 0.00% | 2.16% | 1.94% | 0.28% | 3.34% | 0.20% | 0.15% | 0.44% |
| UK | 0.28% | 0.57% | 0.37% | 0.77% | 0.68% | 1.23% | 42.37% | 12.28% | 0.94% | 41.37% |
| Japan | 0.00% | 0.00% | 0.34% | 1.40% | 5.30% | 7.62% | 8.17% | 11.30% | 27.41% | 21.55% |
| **Emerging** | 0.16% | 0.03% | 0.00% | 0.61% | 0.68% | 0.81% | 9.17% | 3.84% | 17.56% | 27.83% |
| Russia | 0.00% | 0.02% | 0.62% | 1.13% | 1.81% | 1.58% | 3.91% | 12.21% | 8.24% | 6.21% |
| China | 0.05% | 0.01% | 0.00% | 1.18% | 0.85% | 1.59% | 19.66% | 6.66% | 2.50% | 4.45% |
| Brazil | 0.23% | 0.00% | 0.15% | 0.78% | 0.91% | 1.70% | 43.46% | 3.30% | 4.32% | 16.01% |
| **Frontier** | 0.00% | 0.18% | 0.10% | 0.16% | 0.64% | 1.68% | 12.74% | 13.14% | 1.39% | 1.88% |
| Estonia | 0.01% | 0.00% | 0.01% | 0.05% | 0.95% | 0.80% | 8.58% | 1.09% | 0.37% | 0.89% |
| Morocco | 0.13% | 0.01% | 0.00% | 0.08% | 8.04% | 34.85% | 4.93% | 1.82% | 40.05% | 8.23% |
| Jordan | 0.02% | 0.13% | 0.00% | 0.00% | 1.45% | 16.44% | 37.27% | 58.55% | 28.72% | 26.73% |
| **Grand Total** | **0.08%** | **0.08%** | **0.13%** | **0.74%** | **2.07%** | **5.82%** | **16.66%** | **10.40%** | **10.99%** | **16.24%** |

This table reports the contribution (in percentage terms) of RSI rules to the DFDR$^{+/-}$ procedure survivors for every single year examined. For every year, the IS portfolio runs for two years and the remaining eleven are calculated by rolling forward the IS by one month.

**Table B.4**
**Percentage decomposition of filter rules among the DFDR$^{+/-}$ procedure survivors (IS 2 Years)**

| Average of FR % | 2006 | 2007 | 2008 | 2009 | 2010 | 2011 | 2012 | 2013 | 2014 | 2015 |
|---|---|---|---|---|---|---|---|---|---|---|
| **Advanced** | 5.36% | 2.88% | 5.25% | 22.20% | 19.37% | 3.94% | 8.99% | 2.47% | 1.46% | 13.53% |
| US | 50.00% | 17.78% | 8.50% | 5.00% | 7.28% | 2.18% | 10.99% | 2.44% | 2.48% | 1.77% |
| UK | 5.82% | 2.58% | 0.00% | 34.85% | 24.11% | 3.10% | 6.24% | 14.32% | 3.06% | 19.14% |
| Japan | 1.20% | 0.00% | 6.51% | 11.77% | 18.37% | 20.52% | 7.49% | 10.58% | 3.29% | 9.52% |
| **Emerging** | 10.22% | 21.58% | 30.87% | 32.25% | 19.39% | 5.11% | 36.13% | 25.96% | 21.69% | 16.50% |
| Russia | 4.06% | 3.27% | 26.58% | 32.41% | 36.58% | 3.85% | 24.44% | 26.30% | 11.78% | 13.47% |
| China | 3.96% | 5.15% | 5.77% | 11.14% | 2.56% | 3.09% | 13.15% | 17.31% | 15.03% | 9.97% |
| Brazil | 9.04% | 0.51% | 5.07% | 17.07% | 25.09% | 3.24% | 8.86% | 18.34% | 16.25% | 6.16% |
| **Frontier** | 5.30% | 21.67% | 12.69% | 14.68% | 16.85% | 8.15% | 26.43% | 38.02% | 9.49% | 14.65% |
| Estonia | 4.60% | 21.71% | 6.71% | 5.36% | 11.41% | 5.73% | 5.23% | 4.70% | 5.83% | 5.24% |
| Morocco | 8.28% | 2.53% | 1.43% | 2.23% | 9.14% | 8.04% | 0.34% | 9.15% | 8.87% | 11.15% |
| Jordan | 3.82% | 7.21% | 10.79% | 8.76% | 6.14% | 9.69% | 13.30% | 12.82% | 10.03% | 10.02% |
| **Grand Total** | **9.31%** | **8.91%** | **10.01%** | **16.48%** | **16.36%** | **6.39%** | **13.47%** | **15.20%** | **9.11%** | **10.93%** |

This table reports the contribution (in percentage terms) of filter rules to the DFDR$^{+/-}$ procedure survivors for every single year examined. For every year, the IS portfolio runs for two years and the remaining eleven are calculated by rolling forward the IS by one month.

**Table B.5**
**Percentage profitability of moving average rules among the DFDR$^{+/-}$ procedure survivors (IS 2 Years)**

| Average of MA % | 2006 | 2007 | 2008 | 2009 | 2010 | 2011 | 2012 | 2013 | 2014 | 2015 |
|---|---|---|---|---|---|---|---|---|---|---|
| **Advanced** | 71.94% | 82.47% | 78.36% | 57.38% | 50.75% | 88.23% | 23.96% | 90.22% | 92.52% | 22.00% |
| US | 6.25% | 8.89% | 78.16% | 71.88% | 63.52% | 92.80% | 24.34% | 89.08% | 89.08% | 90.59% |
| UK | 81.59% | 84.10% | 93.72% | 35.79% | 41.75% | 86.49% | 34.62% | 41.69% | 88.68% | 19.68% |
| Japan | 89.63% | 51.11% | 68.63% | 42.85% | 18.41% | 21.88% | 35.69% | 26.96% | 12.28% | 23.56% |
| **Emerging** | 67.93% | 40.98% | 49.19% | 31.89% | 51.60% | 86.35% | 20.00% | 25.15% | 31.20% | 25.50% |
| Russia | 91.05% | 91.08% | 55.69% | 40.09% | 31.86% | 87.68% | 20.81% | 17.95% | 21.59% | 27.55% |
| China | 84.00% | 80.75% | 77.00% | 61.10% | 88.94% | 88.69% | 21.92% | 32.10% | 54.53% | 47.02% |
| Brazil | 78.54% | 92.85% | 75.61% | 47.24% | 46.92% | 90.42% | 34.60% | 32.52% | 24.52% | 40.89% |
| **Frontier** | 77.13% | 27.74% | 67.94% | 59.69% | 52.33% | 73.86% | 30.68% | 12.48% | 78.98% | 47.96% |
| Estonia | 78.76% | 32.18% | 74.36% | 77.91% | 61.81% | 82.08% | 50.21% | 81.64% | 77.24% | 77.25% |
| Morocco | 67.01% | 88.31% | 88.22% | 83.70% | 55.28% | 19.57% | 81.08% | 59.27% | 22.58% | 28.34% |
| Jordan | 79.08% | 76.36% | 67.15% | 74.73% | 72.38% | 50.38% | 37.27% | 9.40% | 42.21% | 32.29% |
| **Grand Total** | **72.74%** | **63.07%** | **72.84%** | **57.02%** | **52.96%** | **72.37%** | **34.60%** | **43.21%** | **52.95%** | **40.22%** |

This table reports the contribution (in percentage terms) of moving average rules to the DFDR$^{+/-}$ procedure survivors for every single year examined. For every year, the IS portfolio runs for two years and the remaining eleven are calculated by rolling forward the IS by one month.

**Table B.6**
**Percentage decomposition of support and resistance rules among the DFDR$^{+/-}$ procedure survivors (IS 2 Years)**

| Average of SR % | 2006 | 2007 | 2008 | 2009 | 2010 | 2011 | 2012 | 2013 | 2014 | 2015 |
|---|---|---|---|---|---|---|---|---|---|---|
| **Advanced** | 7.45% | 5.32% | 7.18% | 7.61% | 12.16% | 1.67% | 16.47% | 2.25% | 0.99% | 11.88% |
| US | 31.25% | 4.44% | 5.10% | 12.81% | 10.70% | 0.99% | 31.02% | 2.07% | 1.29% | 1.01% |
| UK | 4.12% | 4.01% | 4.43% | 11.34% | 10.73% | 3.37% | 0.00% | 2.30% | 2.89% | 14.69% |
| Japan | 3.55% | 0.00% | 7.54% | 14.40% | 28.47% | 32.45% | 23.57% | 37.39% | 44.96% | 31.83% |
| **Emerging** | 7.82% | 5.94% | 6.54% | 12.26% | 10.29% | 3.16% | 25.41% | 29.49% | 19.01% | 19.83% |
| Russia | 3.05% | 2.43% | 6.91% | 8.26% | 9.18% | 2.65% | 41.90% | 34.56% | 49.05% | 35.24% |
| China | 3.28% | 3.39% | 8.14% | 11.40% | 2.24% | 3.93% | 44.13% | 38.76% | 22.93% | 29.51% |
| Brazil | 3.28% | 3.19% | 7.70% | 10.90% | 8.50% | 2.74% | 13.08% | 28.94% | 27.41% | 27.34% |
| **Frontier** | 8.10% | 25.65% | 7.16% | 9.67% | 11.94% | 5.84% | 21.87% | 24.96% | 5.15% | 21.97% |
| Estonia | 7.38% | 22.49% | 7.29% | 8.00% | 13.53% | 6.90% | 31.38% | 9.87% | 8.86% | 11.97% |
| Morocco | 6.13% | 2.14% | 3.13% | 9.06% | 20.12% | 34.05% | 13.65% | 21.45% | 24.73% | 49.36% |
| Jordan | 8.43% | 13.61% | 15.47% | 12.17% | 15.34% | 11.92% | 8.43% | 16.67% | 14.53% | 24.94% |
| **Grand Total** | **7.82%** | **7.72%** | **7.22%** | **10.66%** | **12.77%** | **9.14%** | **22.58%** | **20.73%** | **18.48%** | **23.30%** |

This table reports the contribution (in percentage terms) of support and resistance rules to the DFDR$^{+/-}$ procedure survivors for every single year examined. For every year, the IS portfolio runs for two years and the remaining eleven are calculated by rolling forward the IS by one month.

**Table B.7**
**Percentage decomposition of channel breakout rules among the DFDR$^{+/-}$ procedure survivors (IS 2 Years)**

| Average of CB % | 2006 | 2007 | 2008 | 2009 | 2010 | 2011 | 2012 | 2013 | 2014 | 2015 |
|---|---|---|---|---|---|---|---|---|---|---|
| **Advanced** | 15.13% | 9.33% | 9.21% | 12.28% | 16.09% | 4.89% | 44.26% | 4.64% | 4.88% | 13.29% |
| US | 12.50% | 68.89% | 8.25% | 8.15% | 16.57% | 3.75% | 30.32% | 6.21% | 7.01% | 6.19% |
| UK | 8.20% | 8.73% | 1.48% | 17.26% | 22.74% | 5.81% | 16.77% | 29.41% | 4.43% | 5.12% |
| Japan | 5.62% | 48.89% | 16.97% | 29.58% | 29.45% | 17.54% | 25.07% | 13.77% | 12.06% | 13.53% |
| **Emerging** | 13.86% | 31.47% | 13.40% | 22.98% | 18.05% | 4.57% | 9.28% | 15.56% | 10.54% | 10.33% |
| Russia | 1.84% | 3.19% | 10.19% | 18.11% | 20.57% | 4.24% | 8.94% | 8.98% | 9.33% | 17.53% |
| China | 8.71% | 10.70% | 9.09% | 15.18% | 5.41% | 2.71% | 1.13% | 5.18% | 5.01% | 9.05% |
| Brazil | 8.91% | 3.45% | 11.48% | 24.01% | 18.58% | 1.91% | 0.00% | 16.91% | 27.49% | 9.61% |
| **Frontier** | 9.47% | 24.76% | 12.11% | 15.79% | 18.23% | 10.48% | 8.28% | 11.40% | 4.99% | 13.54% |
| Estonia | 9.25% | 23.61% | 11.62% | 8.68% | 12.30% | 4.49% | 4.60% | 2.70% | 7.71% | 4.65% |
| Morocco | 18.44% | 7.01% | 7.22% | 4.92% | 7.42% | 3.49% | 0.00% | 8.30% | 3.76% | 2.93% |
| Jordan | 8.65% | 2.70% | 6.59% | 4.35% | 4.69% | 11.57% | 3.75% | 2.56% | 4.50% | 6.01% |
| **Grand Total** | **10.05%** | **20.23%** | **9.80%** | **15.11%** | **15.84%** | **6.29%** | **12.70%** | **10.47%** | **8.48%** | **9.31%** |

This table reports the contribution (in percentage terms) of channel breakout rules to the DFDR$^{+/-}$ procedure survivors for every single year examined. For every year, the IS portfolio runs for two years and the remaining eleven are calculated by rolling forward the IS by one month.

**Table B.8**
**Annualized Returns and Sharpe Ratios after Transaction Costs (IS 1 Year and OOS 1 Month)**

| Market | 2006 | 2007 | 2008 | 2009 | 2010 | 2011 | 2012 | 2013 | 2014 | 2015 | Average |
|---|---|---|---|---|---|---|---|---|---|---|---|
| **Advanced** | 11.09% (1.27) | -15.38% (-1.66) | 3.43% (0.16) | 3.68% (0.25) | -8.94% (-1.96) | -13.85% (-1.67) | -2.57% (-0.87) | 6.76% (1.09) | -6.95% (-1.77) | -3.07% (-1.15) | **-2.58% (-0.63)** |
| US | 8.51% (0.96) | -13.79% (-1.11) | 5.5% (0.21) | 0.28% (0.02) | -9.76% (-1.47) | -12.72% (-1.55) | -1.17% (-0.35) | 11.66% (1.55) | 0.51% (0.08) | -0.56% (-0.3) | **-1.15% (-0.2)** |
| UK | 16.38% (1.49) | 1.7% (0.17) | 7.01% (0.31) | 19.95% (1.1) | -13.86% (-2.08) | -11.93% (-1.09) | -2.91% (-2.43) | -2.07% (-0.36) | -5.19% (-1.27) | -4.14% (-0.92) | **0.49% (-0.51)** |
| Japan | -10.2% (-0.63) | -4.73% (-0.77) | 19.12% (0.67) | 2.08% (0.09) | -7.74% (-1.41) | -13.7% (-1.64) | -2.77% (-1.62) | -2.29% (-0.19) | -6.09% (-3.11) | -4.72% (-0.91) | **-3.11% (-0.95)** |
| **Emerging** | 3.72% (0.31) | 7.74% (0.44) | 23.23% (0.64) | 0.81% (0.05) | -7.53% (-1.25) | -14.17% (-2.25) | -8.37% (-1.86) | -3.3% (-0.95) | -9.02% (-2.94) | -4.63% (-0.79) | **-1.15% (-0.86)** |
| Russia | 10.33% (0.43) | -8.26% (-0.9) | 49.31% (0.83) | 9.41% (0.36) | -10.31% (-1.04) | -27.52% (-2.51) | -3.66% (-0.51) | -5.97% (-0.9) | 8.68% (0.43) | -15.74% (-1.27) | **0.63% (-0.51)** |
| China | 58.42% (2.72) | 13.24% (0.57) | -0.35% (-0.01) | -13.12% (-0.69) | -10.18% (-1.41) | -19.89% (-2.41) | -2.99% (-1.57) | -6.18% (-0.93) | -8.73% (-1.12) | 11.91% (0.77) | **2.21% (-0.41)** |
| Brazil | -10.06% (-0.43) | 17.71% (0.67) | 84.71% (1.47) | 21.25% (0.98) | -12.27% (-1.62) | -18.28% (-2.17) | -14.19% (-2.37) | -3.53% (-0.63) | -4.38% (-0.36) | 9.01% (0.4) | **7% (-0.41)** |
| **Frontier** | -12.23% (-1.35) | 15.77% (1.53) | 33.12% (1.56) | 11.75% (1.15) | 4.54% (1.1) | -2.12% (-0.36) | -7.06% (-2.01) | -4.42% (-1.16) | 0.68% (0.13) | 9.22% (1.18) | **4.93% (0.18)** |
| Estonia | -18.25% (-1.59) | -8.47% (-0.55) | 68.02% (1.5) | 12.74% (0.52) | 1.8% (0.13) | -6.9% (-0.42) | 1.27% (0.2) | -3.46% (-0.43) | 6.46% (0.67) | -14.41% (-2.55) | **3.88% (-0.25)** |
| Morocco | 19.55% (1.26) | 7.48% (0.6) | 32.43% (1.41) | -4.1% (-0.34) | -5.82% (-1.52) | -6.47% (-1.65) | 2.66% (0.22) | -6.1% (-2.37) | -1.88% (-1.21) | -3.56% (-0.56) | **3.42% (-0.42)** |
| Jordan | -8.63% (-0.57) | 2.5% (0.27) | 39.35% (1.31) | -7.1% (-0.6) | -5.33% (-1.44) | -4.97% (-0.72) | -5.6% (-1.27) | -5.18% (-1.26) | -4.02% (-0.94) | -2.6% (-1.24) | **-0.16% (-0.64)** |
| **Average** | **5.72% (0.32)** | **1.29% (-0.06)** | **30.41% (0.84)** | **4.8% (0.24)** | **-7.12% (-1.16)** | **-12.71% (-1.54)** | **-3.95% (-1.2)** | **-2.01% (-0.55)** | **-2.49% (-0.95)** | **-1.94% (-0.61)** | **1.20% (-0.47)** |

This table reports the average OOS annualized returns and Sharpe ratios of twelve portfolios for one year of IS and one month of OOS after transaction costs (rolling forward by one month). For example, the 11.09% annualized return of the advanced markets (2006) is calculated as the average OOS annualized return of twelve portfolios. The first portfolio's OOS return is calculated over January 2006 using as IS the period 01/01/2005-31/12/2005. The remaining eleven OOS returns are calculated by rolling forward the IS by one month. The same logic applies for the Sharpe ratios. The last column and row present the average performance per market across all years and per year respectively.

**Table B.9**
**Annualized Returns and Sharpe Ratios after Transaction Costs (IS 1 Year and OOS 3 Months)**

| Market | 2006 | 2007 | 2008 | 2009 | 2010 | 2011 | 2012 | 2013 | 2014 | 2015 | Average |
|---|---|---|---|---|---|---|---|---|---|---|---|
| **Advanced** | -2.28% (-0.30) | -13.59% (-1.43) | 11.16% (0.57) | -5.01% (-0.38) | -8.71% (-1.93) | -11.44% (-1.39) | -1.66% (-0.62) | 3.71% (0.66) | -1.28% (-0.36) | -3.35% (-1.41) | **-3.25% (-0.66)** |
| US | 3.83% (0.48) | -13.72% (-1.20) | 19.13% (0.75) | -7.39% (-0.47) | -7.10% (-1.11) | -11.95% (-1.56) | 0.11% (0.04) | 8.93% (1.23) | 2.91% (0.44) | -1.35% (-0.57) | **-0.66% (-0.20)** |
| UK | 16.96% (1.51) | -5.58% (-0.53) | 14.57% (0.52) | 9.43% (0.66) | -8.81% (-1.28) | -16.99% (-1.55) | -2.29% (-1.84) | 1.33% (0.25) | -4.49% (-1.10) | -1.92% (-0.44) | **0.22% (-0.38)** |
| Japan | -18.53% (-1.12) | -0.83% (-0.12) | 24.25% (0.84) | -0.44% (-0.02) | -5.29% (-1.05) | -14.05% (-2.07) | -2.44% (-1.20) | -4.51% (-0.44) | -3.76% (-2.28) | -5.09% (-0.96) | **-3.07% (-0.84)** |
| **Emerging** | -6.89% (-0.60) | -0.68% (-0.04) | 8.65% (0.30) | 3.29% (0.26) | -5.16% (-0.91) | -5.96% (-0.91) | -5.4% (-1.24) | -3.66% (-0.95) | -6.26% (-2.31) | -7.74% (-1.39) | **-2.98% (-0.78)** |
| Russia | 11.32% (0.50) | -19.86% (-1.82) | 50.97% (0.92) | 0.85% (0.04) | -8.58% (-0.92) | -27.03% (-2.69) | -1.60% (-0.24) | -6.99% (-1.05) | 1.75% (0.10) | -14.57% (-1.25) | **-1.37% (-0.64)** |
| China | 31.65% (1.52) | 13.12% (0.55) | -12.97% (-0.51) | -8.78% (-0.52) | -8.99% (-1.42) | -8.86% (-1.30) | -3.07% (-1.73) | -7.99% (-1.26) | -7.77% (-1.10) | 5.67% (0.40) | **-0.80% (-0.54)** |
| Brazil | -14.85% (-0.69) | 24.47% (0.75) | 52.01% (1.21) | 11.24% (0.55) | -7.63% (-0.97) | -9.45% (-1.21) | -10.83% (-1.98) | -0.58% (-0.09) | -4.63% (-0.43) | -1.16% (-0.06) | **3.86% (-0.29)** |
| **Frontier** | -16.7% (-2.16) | 7.56% (0.68) | 29.16% (1.59) | -2.46% (-0.32) | 3.26% (0.80) | -8.84% (-1.66) | -4.11% (-1.10) | -4.45% (-1.40) | -5.22% (-1.28) | 6.75% (0.84) | **0.5% (-0.40)** |
| Estonia | -16.4% (-1.38) | -10.4% (-0.77) | 65.84% (1.54) | 5.71% (0.25) | -2.00% (-0.17) | -12.07% (-0.82) | 4.74% (0.73) | -3.48% (-0.46) | 2.14% (0.24) | -9.68% (-2.18) | **2.44% (-0.30)** |
| Morocco | 6.81% (0.50) | 14.73% (1.12) | 11.77% (0.62) | -7.6% (-0.85) | -6.11% (-1.80) | -6.56% (-1.81) | -2.05% (-0.19) | -4.1% (-2.16) | -2.41% (-1.60) | -2.01% (-0.33) | **0.25% (-0.65)** |
| Jordan | 3.76% (0.27) | 5.43% (0.59) | 22.25% (0.88) | -11.33% (-1.13) | -5.80% (-1.74) | -1.03% (-0.15) | -7.32% (-2.06) | -6.12% (-1.45) | -3.78% (-1.21) | -2.23% (-1.20) | **-0.62% (-0.72)** |
| **Average** | **-0.11% (-0.12)** | **0.05% (-0.18)** | **24.73% (0.77)** | **-1.04% (-0.16)** | **-5.91% (-1.04)** | **-11.19% (-1.43)** | **-2.99% (-0.95)** | **-2.33% (-0.59)** | **-2.73% (-0.91)** | **-3.06% (-0.71)** | **-0.46% (-0.53)** |

This table reports the average OOS annualized returns and Sharpe ratios of four portfolios for IS of one year and OOS of three months after transaction costs (rolling forward by one month). For example, the -2.28% annualized return of the advanced markets (2006) is calculated as the average OOS annualized return of twelve portfolios. The first portfolio's OOS return is calculated over the period 01/01/2006-31/03/2006 using as IS the period 01/01//2005-31/12/2005. The remaining eleven OOS returns are calculated by rolling forward the IS and the OOS by one month. The same logic applies for the Sharpe ratios. The last column and row present the average performance per market across all years and per year respectively.

**Table B.10**
**Annualized Returns and Sharpe Ratios after Transaction Costs (IS 1 Year and OOS 6 Months)**

| Market | 2006 | 2007 | 2008 | 2009 | 2010 | 2011 | 2012 | 2013 | 2014 | 2015 | Average |
|---|---|---|---|---|---|---|---|---|---|---|---|
| **Advanced** | -5.63% (-0.78) | -8.98% (-0.86) | 5.18% (0.31) | -2.7% (-0.25) | -5.56% (-1.30) | -8.33% (-1.14) | -0.57% (-0.23) | 3.08% (0.57) | -1.39% (-0.41) | -2.00% (-0.76) | **-2.69% (-0.49)** |
| US | 2.85% (0.36) | -13.17% (-1.06) | 11.5% (0.53) | -3.28% (-0.25) | -4.6% (-0.77) | -7.73% (-1.10) | 1.19% (0.38) | 6.1% (0.89) | 2.69% (0.39) | -2.28% (-0.88) | **-0.67% (-0.15)** |
| UK | 17.37% (1.50) | -11.36% (-1.00) | 9.35% (0.45) | 7.37% (0.53) | -5.19% (-0.73) | -11.88% (-1.14) | -1.75% (-1.53) | 1.02% (0.21) | -4.33% (-1.14) | -3.64% (-0.81) | **-0.30% (-0.37)** |
| Japan | -19.27% (-1.15) | -5.64% (-0.77) | 18.15% (0.57) | -4.71% (-0.29) | -4.35% (-0.91) | -11.51% (-1.79) | -1.66% (-0.53) | -7.96% (-0.89) | -2.73% (-1.72) | -4.56% (-0.83) | **-4.42% (-0.83)** |
| **Emerging** | -10.01% (-0.94) | 0.18% (0.01) | 15.81% (0.55) | 5.71% (0.47) | -2.45% (-0.45) | -8.8% (-1.30) | -2.79% (-0.59) | -6.14% (-1.48) | -5.29% (-2.08) | -4.36% (-0.73) | **-1.81% (-0.65)** |
| Russia | -1.91% (-0.11) | -22.93% (-2.14) | 82.47% (1.20) | -5.17% (-0.25) | -6.1% (-0.74) | -17.8% (-1.59) | -2.53% (-0.41) | -10.39% (-1.56) | 0.65% (0.04) | -8.66% (-0.71) | **0.76% (-0.63)** |
| China | 26.65% (1.37) | 11.83% (0.52) | -4.14% (-0.15) | -5.06% (-0.32) | -4.39% (-0.74) | -8.11% (-1.30) | -2.50% (-1.57) | -10.32% (-1.58) | 0.65% (0.08) | 0.00% (0.00) | **0.46% (-0.37)** |
| Brazil | -20.72% (-0.96) | 31.88% (0.83) | 24.5% (0.80) | 8.45% (0.44) | -3.76% (-0.48) | -9.81% (-1.45) | -7.86% (-1.47) | -0.85% (-0.13) | -11.6% (-1.07) | -10.01% (-0.54) | **0.02% (-0.4)** |
| **Frontier** | -13.27% (-1.96) | 3.57% (0.34) | 22.57% (1.34) | -3.1% (-0.48) | 2.78% (0.68) | -7.36% (-1.51) | -1.9% (-0.52) | -2.31% (-0.68) | -6.02% (-1.69) | 7.67% (0.95) | **0.26% (-0.35)** |
| Estonia | -14.15% (-1.14) | -5.34% (-0.42) | 37.53% (1.05) | 8.79% (0.40) | -0.14% (-0.01) | -17.09% (-1.28) | 2.31% (0.38) | -4.37% (-0.68) | -4.17% (-0.56) | -8.15% (-1.61) | **-0.48% (-0.39)** |
| Morocco | 1.15% (0.09) | 14.15% (1.06) | -0.90% (-0.06) | -7.89% (-0.97) | -5.22% (-1.49) | -4.44% (-1.19) | -2.55% (-0.26) | -1.92% (-1.01) | -2.92% (-1.74) | -5.03% (-0.82) | **-1.56% (-0.64)** |
| Jordan | 4.05% (0.31) | 3.48% (0.38) | 19.08% (0.84) | -11.84% (-1.27) | -4.89% (-1.33) | -1.70% (-0.26) | -7.22% (-2.06) | -5.10% (-1.05) | -3.28% (-1.27) | -2.66% (-1.41) | **-1.01% (-0.71)** |
| **Average** | **-2.74% (-0.28)** | **-0.19% (-0.26)** | **20.09% (0.62)** | **-1.12% (-0.19)** | **-3.65% (-0.69)** | **-9.55% (-1.26)** | **-2.32% (-0.70)** | **-3.26% (-0.62)** | **-3.14% (-0.93)** | **-3.64% (-0.68)** | **-0.95% (-0.50)** |

This table reports the average OOS annualized returns and Sharpe ratios of four portfolios for IS of one year and OOS of six months after transaction costs (rolling forward by one month). For example, the -5.63% annualized return of the advanced markets (2006) is calculated as the average OOS annualized return of twelve portfolios. The first portfolio's OOS return is calculated over the period 01/01/2006-31/06/2006 using as IS the period 01/01//2005-31/12/2005. The remaining eleven OOS returns are calculated by rolling forward the IS and the OOS by one month. The same logic applies for the Sharpe ratios. The last column and row present the average performance per market across all years and per year respectively.